\documentclass[lettersize,journal]{IEEEtran}
\usepackage{amsmath,amsfonts}
\usepackage{algorithmic}
\usepackage{algorithm}
\usepackage{array}
\usepackage[caption=false,font=normalsize,labelfont=sf,textfont=sf]{subfig}
\usepackage{textcomp}
\usepackage{stfloats}
\usepackage{url}
\usepackage{verbatim}
\usepackage{graphicx}
\usepackage{cite}
\usepackage{xcolor, colortbl}
\usepackage[normalem]{ulem}
\begin{document}

\title{\color{black}Real-time Digital RF Emulation -- II: A Near Memory Custom Accelerator}

\author{M. Mukherjee*, X. Mao*, N. Mizanur Rahman*, C. DeLude*,
J. Driscoll*, S. Sharma, P. Behnam, U. Kamal,\\ J. Woo, D. Kim, S. Khan, J. Tong, J. Seo, P. Sinha, M. Swaminathan, T. Krishna, S. Pande, J. Romberg,\\  and S. Mukhopadhyay (*Equal Contribution Authors)
\thanks{This paper was produced by the IEEE Publication Technology Group. They are in Piscataway, NJ.}
\thanks{Manuscript received Oct, 2023.}
\thanks{The authors are with Georgia Institute of Technology, Atlanta, GA, USA. Corresponding author: saibal.mukhopadhyay@ece.gatech.edu}
\thanks{This work was supported by DARPA \& NIWC Pacific
(N66001-20-C-4001). Any opinions, findings and conclusions are those of the author(s) and do not necessarily reflect the views of DARPA or NIWC Pacific.}}



\maketitle

\begin{abstract}
A near memory hardware accelerator, {\color{black}based on a novel direct path computational model}, for real-time emulation of radio frequency systems is demonstrated. Our evaluation of hardware performance {\color{black}uses} both application-specific integrated circuits (ASIC) and field programmable gate arrays (FPGA) methodologies: 1). The {\color{black}ASIC} testchip{\color{black}implementation}, using TSMC 28nm CMOS, {\color{black}leverages} distributed autonomous control to extract concurrency in compute as well as low latency{\color{black}. It achieves} a $518$ MHz per channel bandwidth in a prototype $4$-node system. The maximum emulation range supported {\color{black}in this paradigm} is $9.5$ km with $0.24$ $\mu$s {\color{black}of} per-sample emulation latency. \textcolor{black}{2). The FPGA-based implementation{\color{black}, evaluated on a Xilinx  ZCU104 board, demonstrates a} $9$-node test case (two Transmitters, one Receiver, and $6$ passive reflectors) with an emulation range of $1.13$ km to $27.3$ km at $215$ MHz bandwidth. }
\end{abstract}


\section{Introduction}
\label{sec:introduction}
\IEEEPARstart{V}{irtual} {\color{black}radio frequency (RF)} emulators {\color{black}have the potential to greatly} reduce the {\color{black} cost of testing} wide-bandwidth RF systems used in radar, electronic warfare, and advanced driver
assistance {\color{black} by offering an alternative to field testing}\cite{barcklow2019radio,matai}.
{\color{black}Since} software simulation of physical models is
orders of magnitude slower than real-time, FPGA-based
emulators {\color{black}were developed in \cite{barcklow2019radio,matai,colosseum1,papenfuss} as a means of accelerating computations}. However,  these existing FPGA testbeds cannot simultaneously achieve {\color{black} the} high computational throughput and low computational latency {\color{black} required to viably test many systems. Coupled with scalability, the three main points of focus in next-generation emulator design are throughput, scalability, and latency; necessitating the development of custom accelerators. We begin with a brief overview of these motivating points and where they arise in RF emulation.}

{\color{black}In terms of throughput, since} real-time RF emulation is performed sample-by-sample, instead of in a batch, the maximum frequency of the hardware determines the maximum sample rate of (digitized) RF signals that can be processed, which in turn determines the maximum emulated RF bandwidth. The existing FPGA-based frameworks are limited to $100$ MHz of bandwidth. 

{\color{black}Basic} RF emulation involves accurate modeling of signal propagation delay, path loss, and {\color{black} radar cross section (RCS)\footnote{{\color{black}In part I of this series we refer to this as the scattering profile, but these are equivalent terms.}} scattering between} objects in a scene at a target RF signal bandwidth. The existing tapped delay (TDL) based computational model, as described in  \cite{barcklow2019radio, matai, colosseum1, papenfuss}, requires $O(M^3)$ computation,  which leads to the $16$ PMAC/s compute throughput for $M=200$ interacting objects at $2$ GHz
 of channel bandwidth. {\color{black} Hence, the development of scalable models is extremely important to limit the computational requirements.}

\begin{figure}
	\centering
	\includegraphics[width = 3.2in]{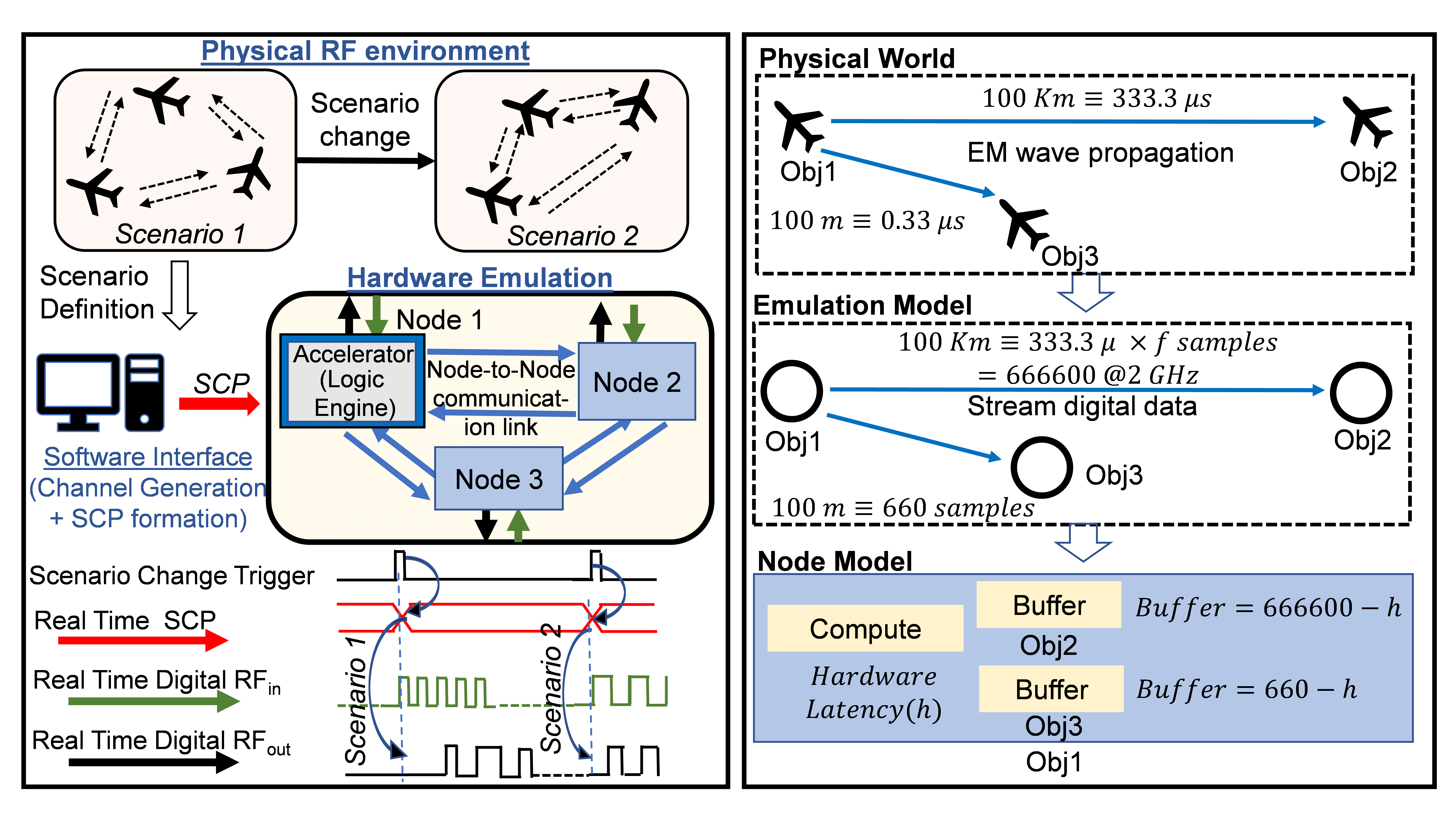}
	\caption{\small \sl Overview of RF emulation system}
	\vspace{-0.2in}
	\label{fig:sys_overview}
\end{figure}

{\color{black}To illustrate the need for low latency}, consider the three-object system in Figure \ref{fig:sys_overview}{\color{black}, where ``Obj\#" denotes ``object number \#"}. {\color{black}Assuming free space propagation} the physical delay of an Electromagnetic (EM) signal from Obj1 to Obj3 ($100$ m spacing) is $0.33$ $\mu$s, whereas, the propagation delay from Obj1 to Obj2 ($100$ km spacing) is $333.3$ $\mu$s. Hence, a custom
accelerator must ensure deterministic processing delay (latency) controlled to always match (time-varying) physical propagation delays between each source and object in the emulated scene. The intrinsic accelerator compute latency must also be    
minimized to {\color{black} reduce the emulator's smallest possible interaction distance.}


{\color{black} In part I of this two-part series, we developed the direct path computational model (DPCM) for RF emulation. This model, which leverages a series of careful factorizations and innovative modeling choices, was shown to be drastically more computationally and memory efficient than its more traditional counterparts. Coupled with an easily distributed computational structure, the model was seen as an attractive option for RF emulator implementation that merited further investigation.}
 
{\color{black}In this paper, part II of the two-part series,} we present the first, to the best of our knowledge, a near-memory accelerator for real-time emulation of interaction among wideband RF signals. {\color{black}It explicitly leverages the DPCM model developed in part I}, reducing the per-sample computational requirement to $O(M^2)$. The compute model is realized using a high-throughput near-memory architecture that leverages concurrent processing to maximize emulated RF bandwidth (chip frequency) as well as minimize the emulated shortest possible distance between objects. 
We design throughput-optimized engines for computing physical models including path loss, RCS, Doppler, and fractional delay correction (FDC)\footnote{{\color{black}A full overview of these phenomena e.g.\ where they arise and how they are modeled, is provided in Part I.}}. A memory-based Single-Input-Multiple-Output FIFO (SIMO-FIFO) employs a novel hybrid control mechanism to simultaneously emulate signal propagation delay to multiple objects. 
The SIMO-FIFO is sub-banked to maximize throughput as well as to ensure the largest physical distance that can be emulated. {\color{black}Distributed local delay distribution} controllers, one connected to each sub-bank, and a deadlock-free multi-casting network-on-chip simultaneously distributes multiple samples per cycle. {\color{black}A global delay distribution} controller configures these distributed controllers to emulate various (and time-varying) physical distances. 

The proposed compute model is validated in software, and the associated architecture is verified via cycle-level simulation for complex scenarios. A $16$-object dynamic RF testcase is simulated on the simulator to validate the architecture. Additionally, a test-chip is fabricated and measured in 28nm CMOS to demonstrate the emulation of a $4$-node system. The test-chip demonstrates range estimation in two different RF scenarios and achieves up to $518$ MHz of RF bandwidth with an emulation range of $0.67$ km - $9.5$ km. 
\textcolor{black}{Finally, to further explore advanced and diverse experimental scenarios beyond the capabilities demonstrated by the ASIC test-chip, we implemented a $9$-node FPGA-based platform. This strategic choice was driven by the desire to extend our research to include a wider array of complex RF signal interaction scenarios. The FPGA platform proves to be exceptionally capable, achieving a bandwidth of up to $215$ MHz and enabling the emulation of RF signal interactions over ranges from $1.13$ km to $27.3$ km.}

The rest of the paper is organized as follows: Section II discusses the related work existing in literature. Section III introduces the RF system, describes the compute model, and discusses the results used to validate it. Section IV provides details on the accelerator architecture (Data Path and Control Path). Section V discusses the verification of the proposed architecture through a C++ cycle-level simulator running various scenarios. \textcolor{black}{Section VI presents the physical design challenges associated with the implementation of the prototype data/control path and the measurement results of experiments performed on the test-chip.} \textcolor{black}{Section VII provides a detailed exposition of the implementation by using an FPGA board, alongside the measured emulated outcomes and the hardware performance metrics of the design.}
The paper is concluded in Section VIII.

\section{Related Work}
\textcolor{black}{In \cite{barcklow2019radio,colosseum1,borries_emulator,buscemi_emulator} the development of hardware based RF emulators is discussed. However, they universally implement an FPGA-based design with the existing TDL model, restricting the maximum allowable RF bandwidth and incurring a higher computational load. Additionally, \cite{borries_emulator} uses separate storage for each real-time sample to be fetched. In \cite{buscemi_emulator}, the possibility of using a single storage buffer is discussed but it results in a large logic-based implementation in FPGA to handle memory contention. 
In \cite{jamin_ims}, a
potential architecture for sparse FIR filtering in RF with
memory-based implementation is explored. The authors discuss a distributed control, co-located
with memory, for sample distribution for sparse FIR filter in
high throughput systems with constrained latency. However,
the work is restricted to architecture-level discussions based on the standard TDL model for RF channels.}

Our prior work in \cite{radarcon, mao} introduces a near-memory-based accelerator architecture for use in RF emulators. Compared to \cite{borries_emulator}, our work uses single storage (for single point object representations) for all outputs for an RF node. In this paper, we expand upon our previous work by adding details in both the compute model and hardware architecture. First, we {\color{black}offer a more detailed discussion} on the hardware {\color{black}architecture, focusing }on the autonomous distributed control that forms the backbone of the {\color{black}design}. Further experiments are also run on the C++ simulator with a 16-object dynamic testcase being presented in this work. {\color{black} Following that, we demonstrate an ASIC-based implementation as a prototype RF emulator for a 4-node system in 28 nm CMOS. We discuss the testchip capabilities and present measurement-based results from testchip experiments to demonstrate the high-performance architecture for emulation of RF interactions. }\textcolor{black}{Finally, we use the ZCU104 FPGA board to build a 9-node system, thereby demonstrating the scalability of our design through further verification of the hardware emulation results. A comparison of our design's performance to state-of-art design is also provided.}

\section{Compute Model}
{\color{black}
The architecture follows the direct path compute model (DPCM) developed in part I of this series. As a brief overview, in this paradigm, each object (containing the accelerator logic engine) acts as both a transmitter and receiver. It constantly sends and receives signals (and applies physical models on the received signals) along the direct path between objects to represent the channels between them. An external control software translates the physical environment to scenario configuration packets (SCPs) and re-configures the accelerator (RF node) every 1 ms (through a scenario programming interface) in response to dynamic changes in emulated scenarios (without stalling real-time operation). For node-to-node communication, high-speed interconnects need to be used for real-time data transfer.

}

\begin{figure}
    \centering
    \subfloat[]{
    \includegraphics[width = 2.2in]{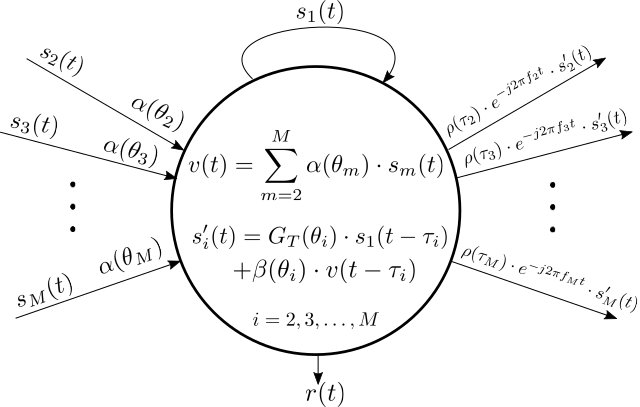}
    }
    \subfloat[]{
    \includegraphics[width = 0.75in]{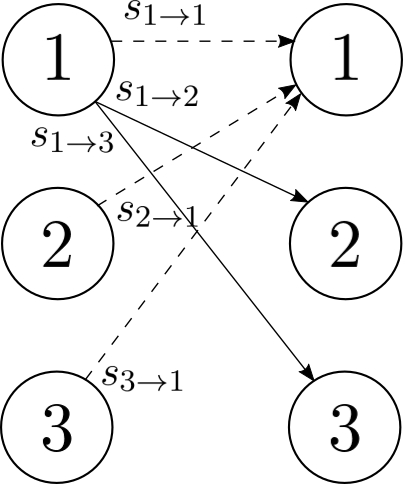}
    }
    \caption{\small \sl DPCM model (a) Computations in DPCM (b) The unrolled communication graph of a simple 3 node DPCM with emitters on the left and receivers on the right. $s_{i\to j}$ denotes a signal sent from node $i$ to node $j$.}
    
    \vspace{-0.2in}
    \label{fig:DPCM}
\end{figure}

{\color{black}A full description of the DPCM model can be found in part I of this series, but we use this section to briefly highlight some of the more major modeling aspects. Though generally congruent with part I, we have made some minor notational changes in this abbreviated description of the model. In particular, we use $M$ to denote the number of nodes and $\rho(\cdot)$ to denote the path loss function, which is now equivalently a function of temporal displacement. Additionally, we use a slightly more verbose labeling of angles and antenna gains. This is simply to help make the variables in the implementation more distinct since there are more of them to keep track of. 

A key feature of the DPCM is its ability to efficiently factor computations. Consider the bistatic scattering response (or RCS) $\sigma(\theta_{in},\theta_{out})$, where $\theta_{in}$ is the incoming angle and $\theta_{out}$ is the outgoing angle. Here we have used the convention that $\theta = (\varphi,\vartheta)$ is a spherical angle with $\varphi$ denoting azimuth and $\vartheta$ denoting elevation. We assume that scattering response admits a separable representation $\sigma(\theta_{in},\theta_{out}) = \alpha(\theta_{in})\beta(\theta_{out})$, which ultimately forms the basis for the factorization. 

Leveraging this separable model we can reduce computations in an $M \times M$ system as depicted in Figure~\ref{fig:DPCM}(a). Each node receives $M-1$ inputs $\{s_m(t)\}_{m=2}^M$ from adjacent nodes at input angles $\{\theta_m\}_{m=2}^M$, where we have arbitrarily reserved $m=1$ to be the node's transmitted signal. These are then weighted by their respective $\alpha(\theta_{m})$ and summed to form an intermediate signal 
\begin{align}
    \label{eq:intermediate_signal}
    v\left(t\right)=\ \sum_{m=2}^{M}{\alpha\left(\theta_{m}\right)s_m(t)}.
\end{align}
This intermediate signal is then buffered in the node and used to generate $M-1$ output signals, where the $i$th output signal is given by
\begin{align}
    \label{eq:output_signal}
    \nonumber
    s_i'(t)& = G_{T}(\theta_i)\cdot s_1(t-\tau_i) + \beta(\theta_i)\cdot v(t-\tau_i)\\
    s_i(t) &= \rho(\tau_i)\cdot e^{-j2\pi f_i t} \cdot s_i'(t).
\end{align}
Here $G_{T}(\cdot)$ is the transmit antenna gain\footnote{\color{black}In this paper we drop the steering angle dependence of the transmit and receive gain functions for brevity, since steering is not explicitly examined in the experiments.}, $f_i$ is the Doppler frequency, and $\tau_i$ is the direct path propagation delay between nodes. During emulation, these parameters, and the function values that depend on them, are updated every SCP. These output signals are then communicated to all adjacent nodes in the network. A simple three-node example of this communication process is presented in Figure~\ref{fig:DPCM}(b).

In addition to the intermediate and output signals given by \eqref{eq:intermediate_signal} and \eqref{eq:output_signal} respectively, a received signal is generated via a separate path. Again we take the $M-1$ input signals and form a weighted combination
\begin{align}
    \label{eq:receive_signal}
    r(t) =\ \sum_{m=2}^{M}{G_R\left(\theta_{m}\right)s_m(t)}.
\end{align}
where $G_{R}(\cdot)$ is the receive antenna gain. \eqref{eq:receive_signal} represents the signal seen by the receiver of an object if it possesses one\footnote{\color{black}For the single scatterer case examined in this paper, we assume the receiver is located at the phase center of the object.}.

}

\begin{figure*}
	\centering
	\includegraphics[width = \textwidth]{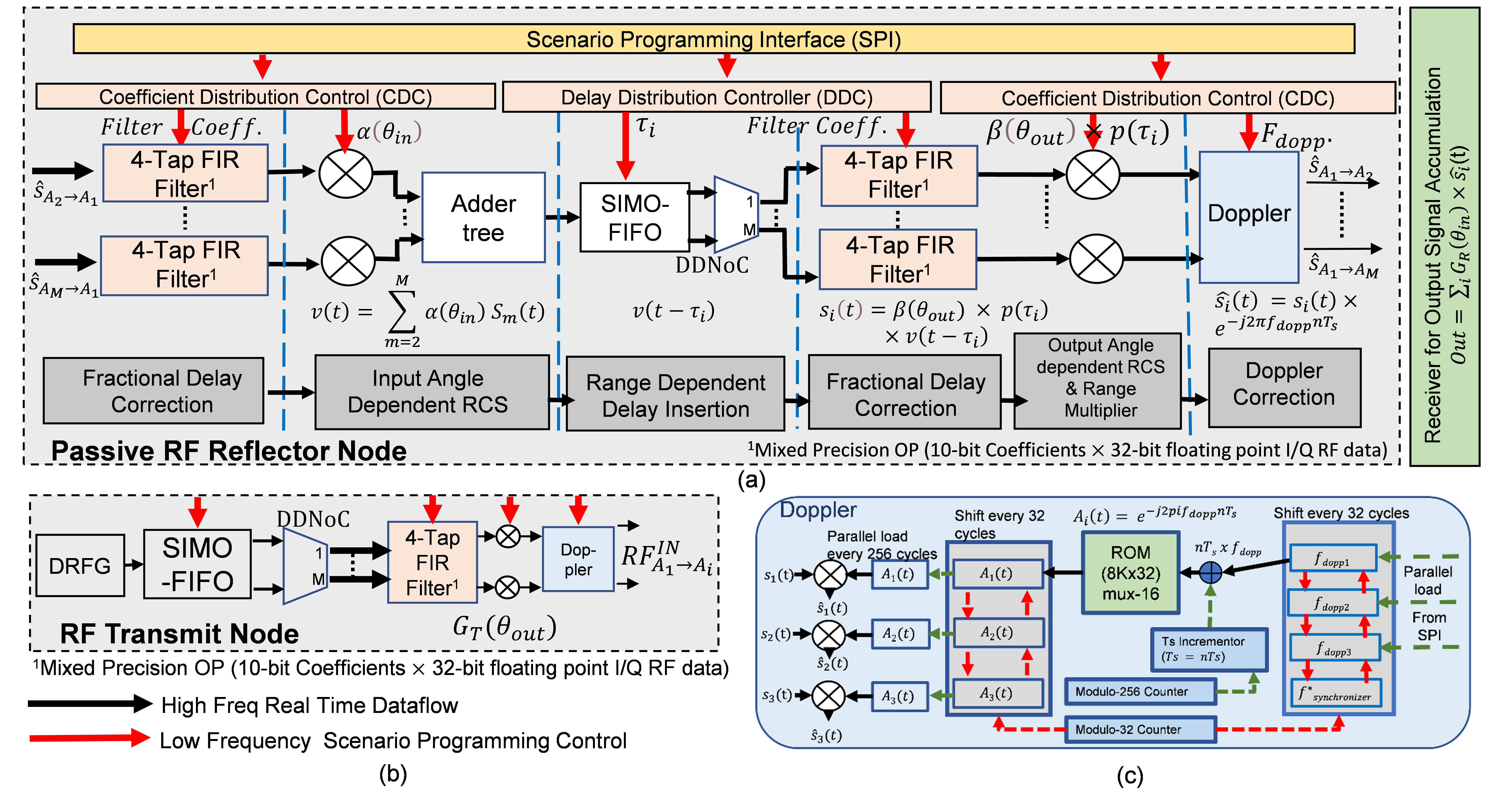}
		\vspace{-0.2in}
	\caption{\small \sl Accelerator architecture based on Direct Path Compute Model (a) Architecture of passive node (b) Architecture of transmit Node (c) Doppler Module}
	\vspace{-0.2in}
	\label{fig:archi}
\end{figure*}
\section{Accelerator Architecture}
In order to faithfully implement the DPCM in hardware, the architecture of a node must emulate several physical phenomena. These phenomena and an accompanying high-level summary of our approach to implementing their respective models are given in Table~\ref{tab:node_phen}. The architecture (Figure~\ref{fig:archi}) consists of the data path (containing all compute modules) and control path (FIFO for data buffering according to direct path delay).
{\color{black}
\begin{table}[t]
    \centering
    \caption{\small \sl Phenomena modeled by each node and the approach taken to implement the model within our architecture.}
    \begin{tabular}{|p{0.14\textwidth}|p{.26\textwidth}|}
        \hline
        \textbf{Phenomenon} & \textbf{Approach} \\
        \hline
         Direct path delay & On-chip memory to buffer and source samples.\\
         \hline
         Doppler & Narrowband Doppler with an update rate of 2 MHz. Resource sharing is used to reduce hardware costs.\\
        \hline
         Fractional delay filter & 4-tap filter running at 25\% oversampling ($\sim 2.5-2.6$ GHz) applied to each object to consider fractional delay between objects' relative phase centers (true broadband processing \& oversampling reduces filter distortion with more accurate delay approximation).\\
        \hline
         Antenna gains & Multiplication in source and/or receiver signals.\\
          \hline
         RCS \& path loss &  Seperable RCS coefficients. Multipliers are used to apply RCS and path loss in architecture.\\
         \hline
    \end{tabular}
    \label{tab:node_phen}
\end{table}
}

\subsection{Data Path}
\textcolor{black}{Figure~\ref{fig:archi}(a) shows the architecture of the hardware accelerator for a passive reflector node.} {\color{black}The RCS of an object is emulated using a point scatterer with respective input and output dependent gains $\{\alpha(\theta_{in}),\beta(\theta_{out})\}$\cite{kscatt}}. Distance dependent direct path loss $\rho(\tau_i)$, angle-dependent 
receive $G_R(\theta_{in})$ antenna gains, and 4-tap FIR filter-based FDCs are used in emulation, all of which are implemented using high throughput multiply and accumulate (MAC) units. $\beta(\theta_{out})$ and $\rho(\tau_i)$ are lumped into a single coefficient $=\beta(\theta_{out}) \times \rho(\tau_i)$) so that only a single multiply unit is needed.

\textcolor{black}{At the final output response generated by an object, independent narrowband Doppler correction is also used on each output to account for relative motions of different objects (Figuure~\ref{fig:archi}(c)). This is implemented using a high throughput complex MAC unit.} The complex Doppler correction coefficient is updated at a reduced frequency of $2$ MHz. This allows a single Doppler coefficient generation block to be shared across all outputs to minimize area and power overhead, taking advantage of the fact that Doppler coefficients need not be updated every cycle. Given an input Doppler frequency, a Finite State Machine (FSM) sequentially generates the complex exponential for Doppler correction for each output (in a round-robin fashion) using a ROM-based lookup table (for optimal area/power) to generate the complex exponential. \textcolor{black}{Each newly generated coefficient is buffered until all of them have been generated. Thereafter, all output Doppler coefficients are simultaneously updated once every $256$ cycles, to allow a single cycle update of Doppler coefficients. The latency of generating a single Doppler coefficient is $32$ cycles and therefore this FSM unit can generate Doppler coefficients for up to $4$ separate outputs to maintain a Doppler update rate of $2$ MHz. For $>4$  outputs, either a second Doppler FSM unit will be required (to maintain the same Doppler update rate) or the Doppler update rate must be reduced if a single Doppler FSM unit is used.}

\textcolor{black}{The receiver associated with each node sums up the signals received at each node after weighing them with the corresponding angle dependent receive gains $G_R(\theta_{in})$.}
\begin{figure*}
	\centering
	\includegraphics[width = \textwidth]{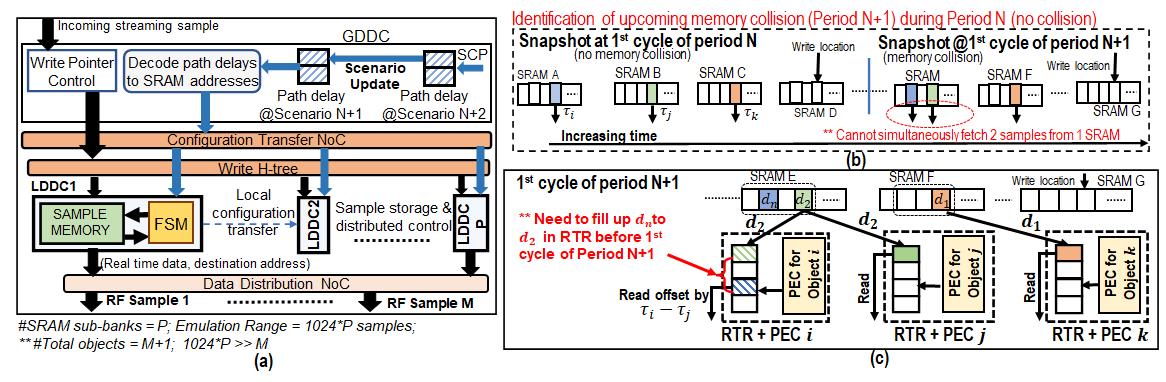}
		\vspace{-0.3in}
	\caption{\small \sl Detailed architecture of sample distribution (a)Sample storage and distribution by LDDC (b)Memory collision (c) Memory collision handling}
	\vspace{-0.2in}
	\label{fig:lddc}
\end{figure*}

\subsection{Control Path}
The main function of the control path is to buffer the scatterer response $v(t)$ in SIMO-FIFO (Figure~\ref{fig:archi}).
For each output object at a path delay $\tau_i$, SIMO-FIFO distributes the appropriately delayed $v(t - \tau_i)$ signals.  
The architecture uses distributed, autonomous control to extract high throughput by maintaining a single input (write) pointer and multiple parallel output (read) pointers (one for each object). The control path contains a Global Delay Distribution Controller (GDDC) and multiple Local Delay Distribution Controllers (LDDC). 

The path delays for objects are realized by controlling the difference between (single) write and (multiple) read pointers. 
Given a physical distance $d$ to be emulated for an RF channel with bandwidth $f$, the SCP determines ``physical delay" (in terms of samples) as $d \times f/c$, where $c$ is the speed of light. The design has a known deterministic processing latency in the signal emulation path. The global
delay distribution controller (GDDC) computes `buffer delay' i.e., the number of cycles a sample must be held in SIMO-FIFO, as $\text{buffer delay }=d \times f/c - \text{compute latency}$ to match the real-world propagation delay. The SIMO-FIFO is implemented as distributed SRAM
modules, 
driving a multiplexer-based data distribution network-on-chip (DDNoC) to send RF samples to the compute engines. For every scenario, the GDDC parses the SCP and
configures distributed Local DDC (LDDC) present with each
SRAM. \textcolor{black}{The Configuration Transfer NoC is used to transfer these configurations from the GDDC to the corresponding LDDC.} The LDDCs are distributed FSMs, maintaining their own read (output) pointers in real-time (Figure~\ref{fig:lddc}(a) shows P LDDCs generating M outputs). 
\textcolor{black}{Considering the write pointer location as the reference point (location of the RF node executing the computation), the output pointer reads data from a location offset by this buffer delay. The write pointer is incremented every cycle to store incoming streaming RF samples in successive memory locations and the read pointers are incremented simultaneously, keeping the delay difference constant (according to the direct path delay of the objects in the system with respect to the computing node).}
The sample to be distributed is first ``buffered" in the SIMO-FIFO according to the computed ``buffer delay." Samples are ``virtually shifted" in the SRAM-based SIMO-FIFO through the use of dynamic pointers which update their location every cycle while the actual data remains ``physically" fixed. This approach allows potential savings in area/power through the use of SRAMs for sample storage. 
The starting sample fetch location for a scenario is decoded based on the buffer delay and the read (output) pointer is sequentially incremented in subsequent cycles during a scenario. \textcolor{black}{The incoming real-time samples are written to the sub-banked SRAMs through throughput optimized on-chip H-tree like network and the samples read by the active LDDCs are distributed to the subsequent compute modules with a data distribution NoC (DDNoC) with P inputs and M outputs.} Once the last sample in an SRAM sub-bank has been accessed, control is transferred within a cycle to the next logically neighboring LDDC to start data access from its associated SRAM.

\begin{figure}
	\centering
	\includegraphics[width = 3.2in]{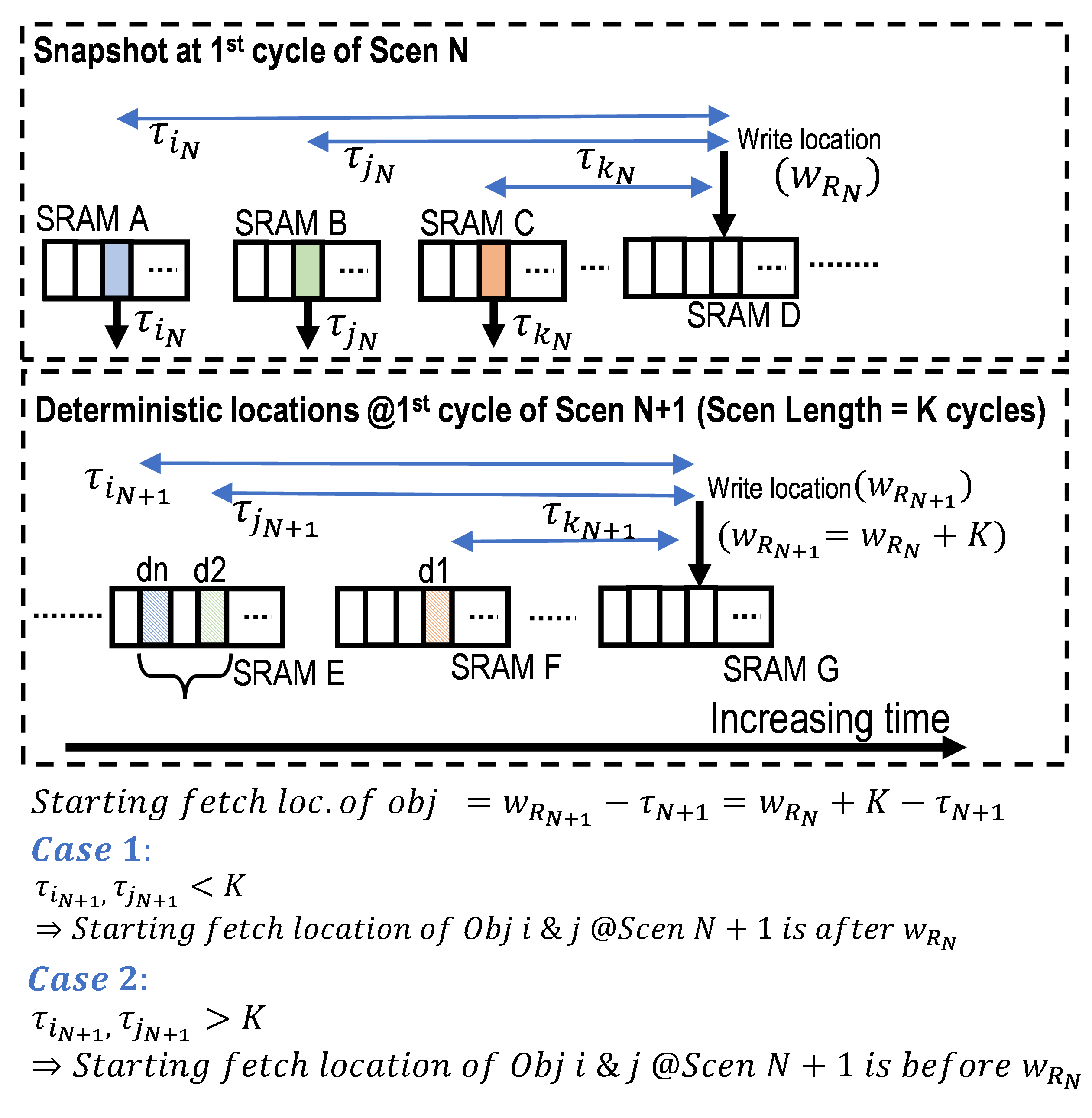}
		\vspace{-0.1in}
	\caption{\textcolor{black}{\small \sl Illustration of data prefetch}}
	\vspace{-0.25in}
	\label{fig:prefetch}
\end{figure}
\subsection{Memory Collision}

Memory read-read conflict (Memory collision)
occurs when ranges of multiple objects are closer than the number of samples stored in a single SRAM. Prior work \cite{buscemi_emulator} discusses the possibility of using a single memory to implement tap delays for multiple objects (for RF emulation with a TDL model) but incur large logic usage in FPGA implementation to mitigate memory contention when the sample to be read for two tap delays are within the same memory sub-bank. 
The memory space representation of a system of 4 objects with memory collision is shown in Figure~\ref{fig:lddc}(b), across two successive scenarios (Scen. N and Scen. N+1). In a cycle, SRAMs allow reading from only one row. Considering 1 data per row, for an SRAM storing S samples, this sets a lower bound on the range difference between objects ($|\tau_i - \tau_j| > S$). $Memory\ collison$ occurs when this condition is violated and multiple samples need to be read from an SRAM in a cycle. 

We use architectural techniques to support memory collision handling by adding real-time registers (RTR), prefetch buffers (PB), and processing engine controllers (PEC) to our control path. For an $M+1$ object system, there are $M$ RTRs and $M$ PBs in one node. The GDDC contains a period look-ahead and an upcoming $memory\ collision$ in Scen. N+1 is known at the beginning of Scen. N. ``Colliding" objects are grouped together (starting from the nearest object) within GDDC. Only the object with the smallest path delay in the group (nearest object in the group), referred to as the ``group header," is parsed by the GDDC to configure the corresponding LDDC for data fetch. However, this data is multicast from the single LDDC to the RTRs (Figure~\ref{fig:lddc}(c)) corresponding to each object in the group. The DDNoC supports arbitrary uni- and multi-casting of multiple (equal to number of output objects) samples. Real-time samples for subsequent compute modules are tapped from the RTRs (supporting both read/write simultaneously). The group-header data ($d_1, d_2$) are read from the most recently updated locations in their RTRs. Read locations for non-group header objects $\tau_i$ are offset from the most recent location by $\tau_i - \tau_j$. Each PEC for an object is associated with an RTR and PB. Similar to LDDCs, they are autonomous FSMs with their own read and write pointers.
Data multicast starts from $d_2$ in Scen. N+1 (Figure~\ref{fig:lddc}(c)), whereas the 1st correct data for object $i$ in Scen. N+1 is $d_n$. Data ($d_n-d_2$) for object $i$ needs to be supplied separately. 
Since object locations are static during a scenario and GDDC contains a scenario look-ahead, 
data $d_n-d_2$ is $prefetched$ in Period N (Figure~\ref{fig:prefetch})  
and stored in PBs present with each RTR. \textcolor{black}{Based on the path delays of objects in a group, the data to be prefetched for Scenario N+1 may or may not be available in the SIMO-FIFO at the start of Scenario N.} \textcolor{black}{Figure~\ref{fig:prefetch} illustrates the two possible prefetch conditions. When the path delay of the objects in a group $>$ scenario length, the data to be prefetched is already present in the sample storage SRAMs at the beginning of Scenario N and the GDDC configures the corresponding LDDC to fetch this prefetch data when it is idle. However, if the path delay of the objects in a group $<$ scenario length, the data to be prefetched will be streaming in during Scenario N. The GDDC tracks this and directly sends this data to the corresponding PBs along with storing it in the SIMO-FIFO.}

\begin{figure*}
	\centering
	\includegraphics[width = 0.9\textwidth]{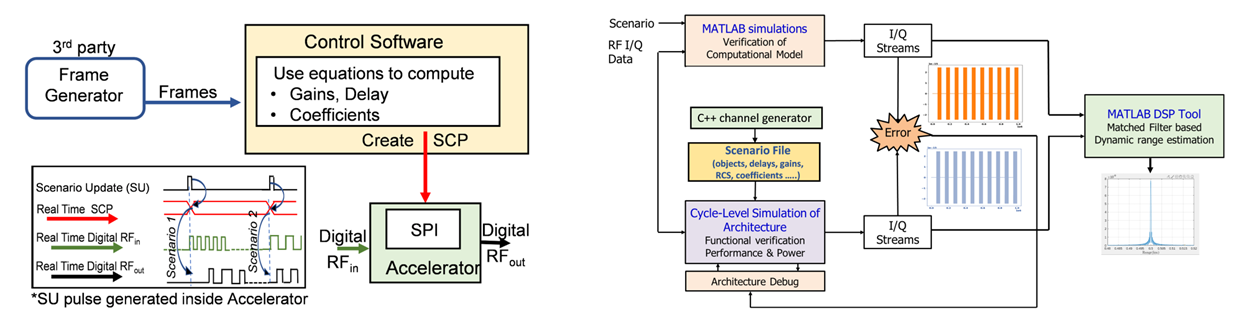}
		\vspace{-0.1in}
	\caption{Emulator Details (a) Scenario Programming (b) C++ cycle level simulator platform}
	\vspace{-0.1in}
	\label{fig:scen_prog}
\end{figure*}

\begin{figure*}
	\centering
	\includegraphics[width = 0.9\textwidth]{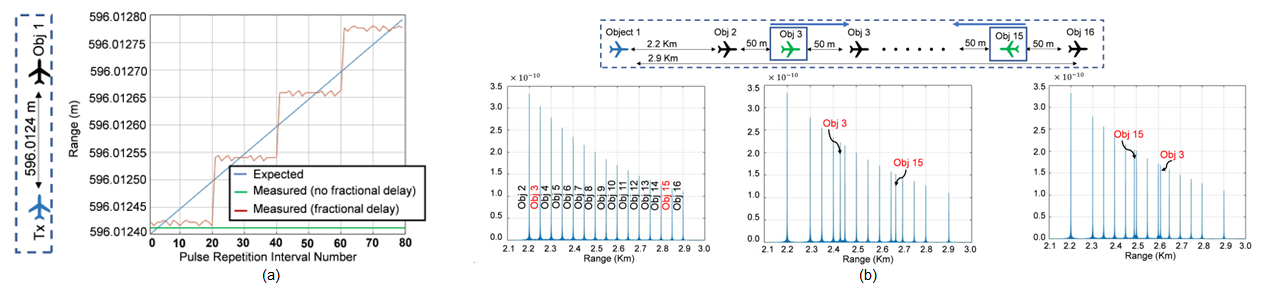}
		\vspace{-0.15in}
	\caption{Architecture verification with C++ simulator (a)Impact of fractional delay connection \textcolor{black}{(b)Large scale 16-object dynamic scenario on the C++ simulator} }
	\vspace{-0.15in}
	\label{fig:large_scen}
\end{figure*}

\subsection{Scenario Programming}

The scenario programming takes a real-life radar scenario and computes model parameters needed by the accelerator (Figure~\ref{fig:scen_prog}(a)). It contains two parts: the frame generator and the control software. The frame generator extracts the scenario information to create messages and frames. The messages provide information, like the number of objects, about the scenario to set up the control software for the frames. The frames describe each object's properties like position, velocity, acceleration, and quaternion orientation and they are updated every millisecond. Both the messages and frames are sent to the control software through a shared memory.  

\textcolor{black}{On the other end, the control software is a multi-threaded process that solves respective equations to solve for parameters (Antenna gains, Doppler frequency, propagation delays, etc.). It runs with 8-32 threads based on the number of objects in the scenario with three threads to read frames and messages from the shared memory and one main thread to control the execution of the software. The rest of the threads are used to calculate the parameters by distributing the calculations between the threads so a thread can calculate the parameters for eight objects.} Since scenario updates are applied every 1 ms, each frame’s equations get solved within $1$ ms by vectorizing the equations’ implementation using Intel intrinsic, which are C-style functions to access specific ISA instructions without writing assembly code. For example, one thread can do the same add to all eight objects with one ISA instruction. The output parameters corresponding to each frame (SCPs) are transferred to the accelerator through a Scenario Programming Interface (SPI).

Within the accelerator, gains are double buffered and loaded with an internally generated scenario update (SU) pulse, aligned to the negative edge of global CLK to prevent stalling of RF emulation for configuring new scenarios.
The SCPs are double-buffered inside the GDDC and parsed a scenario ahead to send configuration information to LDDCs.

\section{Verification of Proposed Architecture}
\begin{figure}
	\centering
	\includegraphics[width = 2.6in]{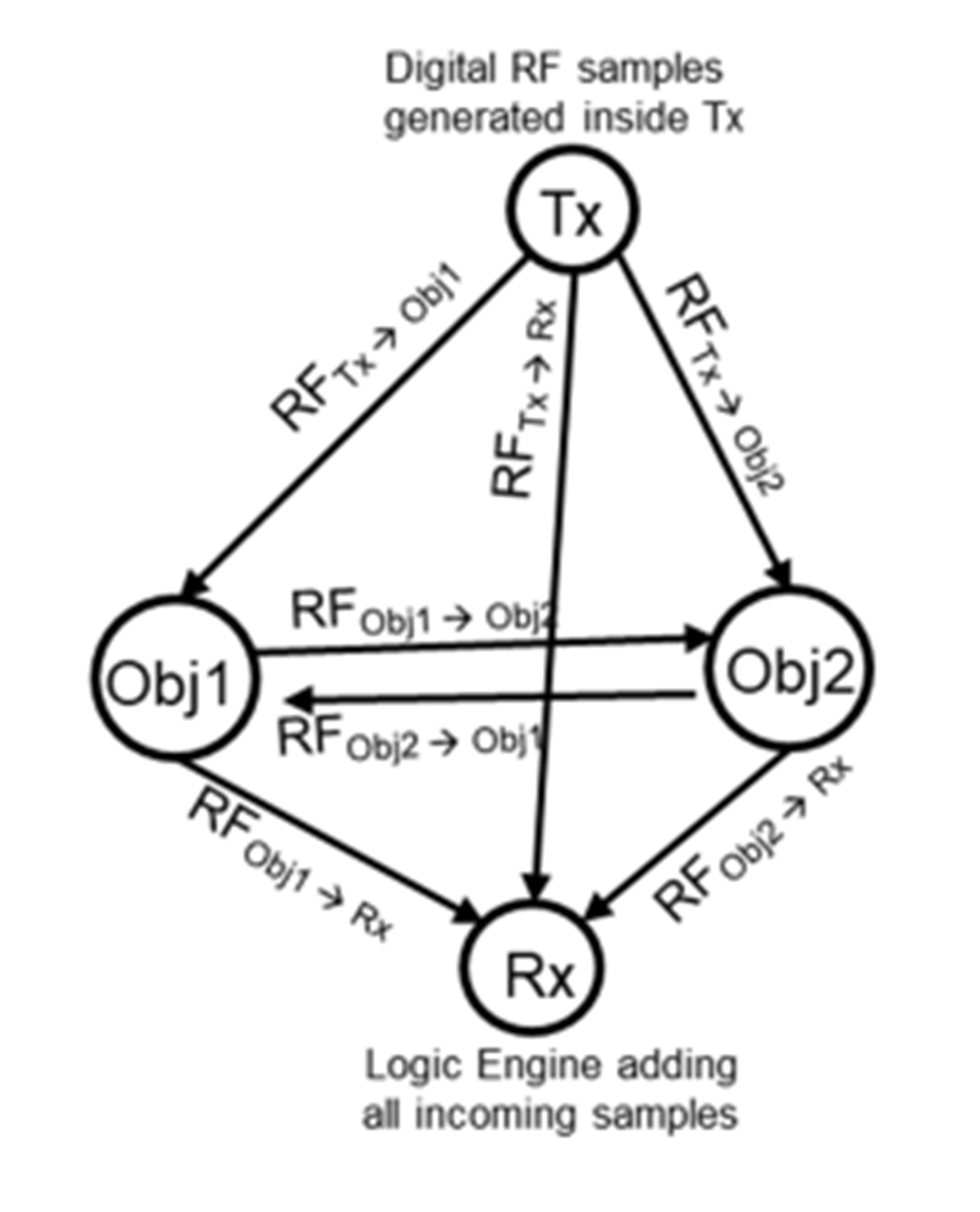}
		\vspace{-0.3in}
	\caption{\textcolor{black}{\small \sl RF Application Subsystem on test-chip for demonstration }}
	\vspace{-0.15in}
	\label{fig:chip_implementation}
\end{figure}
\begin{table}[h]
    \centering
    \caption{\small \sl Bit precision exploration of FDC coefficients with full-precision data.}
    \begin{tabular}{|p{0.1\textwidth}|p{.1\textwidth}|p{.1\textwidth}|}
        \hline
        \textbf{Bit-precision of coefficients (Floating point)} & \textbf{Computed sample (Accuracy)} & \textbf{Range resolution (Accuracy)}\\
        \hline
        16-bit & 0.2075 (100\%) & 0.0461 (100\%)\\
        \hline
        14-bit & 0.2076 & 0.0462\\
        \hline
        12-bit & 0.2073 & 0.0462\\
        \hline
        11-bit & 0.2091 & 0.0435\\
        \hline
        10-bit & 0.2093 & 0.0445\\
        \hline
    \end{tabular}
    \label{tab:bit_prec}
\end{table}

\subsection{Cycle Level Simulator}
We develop a sample and cycle-level simulator in C++ to evaluate the performance of the proposed architecture in emulating RF interactions (Figure~\ref{fig:scen_prog}(b)). For a specific RF scenario, the channel generator generates ``Scenario File" containing the compute model parameters such as path gains, RCS, direct path delay, etc. These files are used as input by the SPI of the C++ simulator to send scenario-specific control information (SCPs) to the necessary modules. Matched filtering of the output with the input chirp signal is used to determine the range of the object. The I/Q streams from MATLAB computation and C++ platform-based emulation are compared to validate the architecture.

\subsection{Testcases on cycle level simulator}

\subsubsection{Testcase 1}
This testcase demonstrates the importance of fractional delay filters proposed in our architecture. At $2.5$ GHz operating frequency, the emulated range resolution is $\approx 0.12m$ per sample. Since we update the direct path delay coefficients after each frame, the emulation can not capture (without introducing fractional delay) change in a range less than $0.12$ m per frame ($1$ ms in our case) which can be a problem, especially for a very slow-moving object. Therefore, we consider a simple testcase with 2 objects where Object 1 is located at $596.012$ m distance away from the Tx (at origin) moving at a velocity of $0.1$ m/s.  
The scenario update occurs after every $20$-pulse repetition interval (PRI). It is evident that with fractional delay filters, the estimated range resolution improves significantly (Figure~\ref{fig:large_scen}(a)).

\subsubsection{Testcase 2}
\textcolor{black}{In this simulation we present a large $16$-object dynamic scenario with memory collision to demonstrate the handling of complex RF scenarios with the proposed architecture (Figure~\ref{fig:large_scen}(b)). Object 1 is located at the origin and the range estimation of Object 2 to Object 16 is carried out over a series of scenarios. Objects 2 to Object 16 are located at intervals of $50$ m starting from a distance of $2.2$ km from Object 1 (Scenario 1). This is a testcase with memory collision since the distance between objects is $50$ m which translates to a $\sim 400$ sample delay (The number of samples stored in each SRAM was set at 1024). Object 3 and Object 15 have relative motion compared to Object 1. Object 3 starts from the range of $2.25$ km and moves away from Object 1 whereas Object 15 starts from $2.85$ km away from Object 1 and gradually moves towards it. The results show the estimated ranges across scenarios and it clearly shows that the ranges of all objects have been detected in the presence of memory collision and that objects 3 and 15 are moving. Additionally, the amplitude of the detected peaks get smaller as the range of the detected object increases, indicating the modeling of physical phenomena by the simulated architecture. }

\section{Test-chip Implementation} 

\begin{figure}
	\centering
	\includegraphics[width = 2.8in]{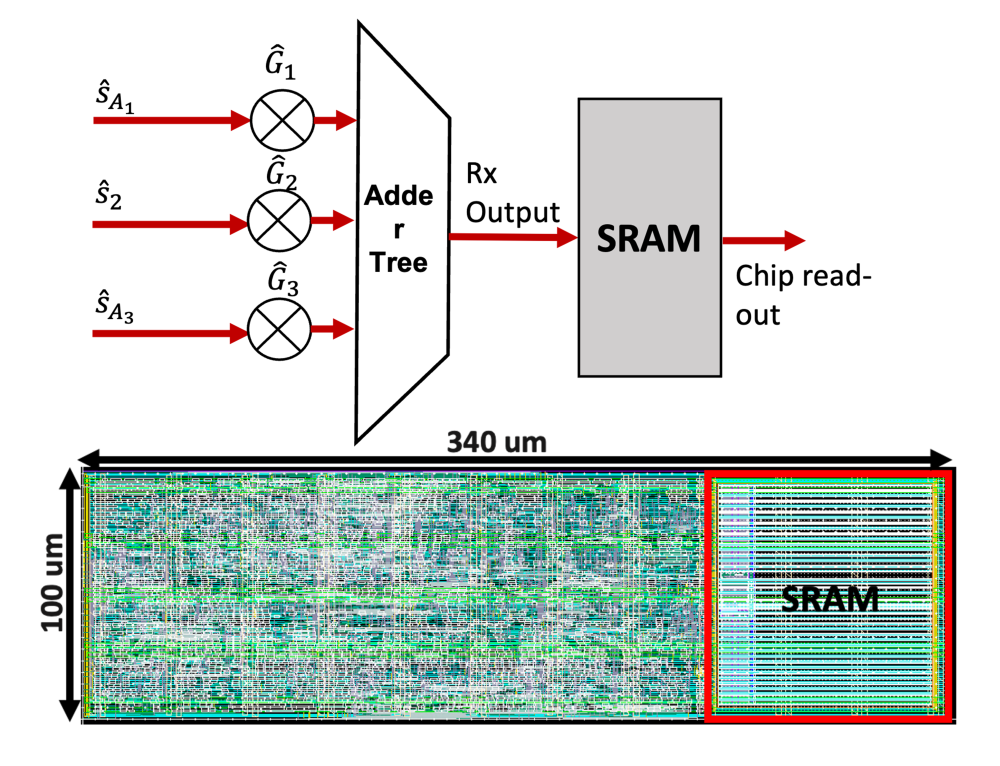}
		\vspace{-0.15in}
	\caption{\textcolor{black}{\small \sl Testchip implementation details of Receiver }}
	\vspace{-0.1in}
	\label{fig:RX}
\end{figure}

\begin{figure*}
	\centering
	\includegraphics[width = \textwidth]{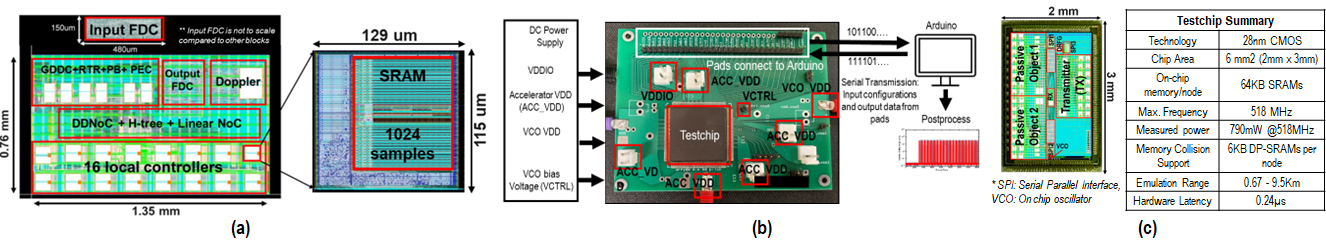}
		\vspace{-0.15in}
	\caption{\textcolor{black}{\small \sl Prototype summary (a)Layout of a single passive node with the layout of LDDC+SRAM (b)Experimental Setup (c)Die-shot and testchip summary }}
	\vspace{-0.15in}
	\label{fig:chip}
\end{figure*}

A prototype emulator is fabricated in $28$ nm CMOS with Synthesis and Place/Route tools. We choose a real-time RF system with 4 interacting nodes as an application to demonstrate the distributed control (Figure~\ref{fig:chip_implementation}). 
The system contains a Transmitter, 2 reflectors (Objects 1, 2) and a Receiver. The Tx generates high throughput digital samples which are reflected at/between passive objects and collected at the Rx.

\subsection{Implementation Details of Data Path}
Coefficients $\alpha(\theta_{in})$ and $\beta(\theta_{out})$ are implemented using 16-bit floating point multipliers (both data and coefficients are 16-bits). Each complex Doppler coefficient is also implemented using 16-bit precision MAC units to preserve accuracy as the impact of these MAC units on overall area and power is minimal due to their small number per node ($2\times$ the number of outputs for complex coefficients). However, due to the large number of total MAC units required for FDC ($8\times$ the number of inputs/outputs), these 4-tap FIR filter-based FDCs are implemented as a mixed precision MAC unit with reduced precision (5-bit exponent + 4-bit mantissa + 1-bit sign = 10-bits) FIR coefficients and 16-bit data for reduced area/power with minimal impact on accuracy. This mixed precision MAC unit provides a $20\%$ and $25\%$ reduction in power and area respectively compared to a 16-bit coefficient MAC unit while incurring a $0.87\%$ reduction in accuracy of computed sample and $3.47\%$ reduction in range resolution (Table~\ref{tab:bit_prec}).
\textcolor{black}{For a transmitting node, the first part of the data path (until the generation of intermediate signal $v(t)$) is replaced by a local RF source.} The digital RF generator (DRFG) is emulated using a configurable FSM, generating digitized 32-bit I/Q (16-bit I + 16-bit Q) RF data (Figure~\ref{fig:archi}(b)). \textcolor{black}{The digital output I/Q stream is a periodic signal with programmable time period, duty cycle, and I/Q values that provides flexibility in generating a large variety of RF input signals.} The angle-dependent transmit gain $G_T(\theta_{out})$ is accounted for with the help of 16-bit MAC units in the transmit path. 

\textcolor{black}{The receiver (Figure~\ref{fig:RX}) performs weighted accumulation of data from the Transmit node and the 2 passive nodes. The receiver gains are applied to each input using real-valued 16-bit coefficient MAC units before being combined in an adder tree. For the prototype test-chip implementation, a high-bandwidth I/O interface for real-time readout of the receiver output at operating frequency was not implemented. Instead, the final samples generated at the output of the adder tree, are sequentially stored in a local SRAM for the duration of the emulation. Once the emulation is completed, the stored receiver samples are read out from the SRAM through a low bandwidth on-chip serial to parallel Interface.}

\subsection{Implementation Details of Control Path}

\textcolor{black}{We use a small-scale design of the proposed near-memory distributed control. Synthesis and automatic place \& route tools are used to implement the distributed control with SRAM sub-banking to simultaneously optimize area and performance (while avoiding register-based implementation). This prototype emulator is implemented with the storage of 16 kilo-samples per node (16 SRAMs per node, 1024 samples in each foundry-provided SRAM sub-bank) for demonstration. Each sample is 32-bit wide.} 
\textcolor{black}{Memory collision is handled by the incorporation of RTR/PB in the distributed control. The RTR/PB each needs to store 1024 samples to provide collision support across a sample span of 1024 (full support). 
Similar to sample storage in baseline distributed control, the RTR is implemented with SRAMs, but with dual ports (DP) to facilitate the Read/Write in the same cycle. DP-SRAMs are significantly slower and larger (1 kB memory is $2\times$ larger, $23\%$ slower) than single port SRAMs. We choose DP-SRAMs with 256 samples (1KB) each to simultaneously optimize area/frequency in this test-chip due to the constraints on the available chip area. This restricts the memory collision protection range to 256 samples (1 kB DP-SRAM) instead of the full 1024. 
Parallel loading of multiple prefetch data is not possible in a single cycle in DP-SRAM-based RTR from PB at the end of a Scenario. So, we implement PBs with DP-SRAM too and the roles of RTR and PB are interchanged in every scenario (virtual parallel loading of prefetch data) by PEC logic. Each PE controller is associated with a single 1KB DP-SRAM for RTR and another single 1KB DP-SRAM for PB. The PEC FSM is implemented to increment both input/output pointers every cycle. 
The addition of DP-SRAMs/PECs causes a significant overhead in the area/power of the control path. 
}

\textcolor{black}{From Figure~\ref{fig:archi}(b), the SIMO-FIFO-based sample distribution has the same requirement in a transmitting vs a passive node, with the key difference being that in the Transmit node, the sample from the DRFG is buffered instead of the intermediate signal $v(t)$. Additionally, with the choice of prototype RF subsystem, the Tx sends samples to 3 other nodes while the two passive reflectors send samples to 2 other nodes. We use the same SIMO-FIFO in all 3 nodes, with passive nodes configured to generate only 2 output samples (1 RTR + PB + PEC will be idle).}

\textcolor{black}{The LDDC+SRAM is designed together and used as hard modules in the design of control path. The rest of the control path is created together along with hard modules for the output Fractional Delay correction and Doppler modules to allow the tool maximum flexibility in area/performance optimization. Neighboring LDDC+SRAMs are placed close together (in a ring-like arrangement where the first SRAM and last SRAM sub-bank being logically adjacent are placed next to each other) to facilitate single cycle local transfer of configurations once each LDDC has read the last sample from its associated SRAM.}

\textcolor{black}{Figure~\ref{fig:chip}(a) shows the layout of a single passive node with the placement of the 16 LDDC + SRAM, DP-SRAMs, FDC and Doppler highlighted. The layout of an individual LDDC + SRAM combination is also highlighted in Figure~\ref{fig:chip}(a). The area occupied by a single LDDC + SRAM is $0.014$ $mm^2$ with the 4 kB SRAM occupying $56\%$ of the total area. Roughly, the area occupied by the GDDC and DP-SRAMs is around $0.2$ $mm^2$ (area of dual port SRAMs = $0.04$ $mm^2$ ). The total area occupied by a single passive node is about $1.098$ $mm^2$.  }

\begin{figure*}
	\centering
	\includegraphics[width = 0.85\textwidth]{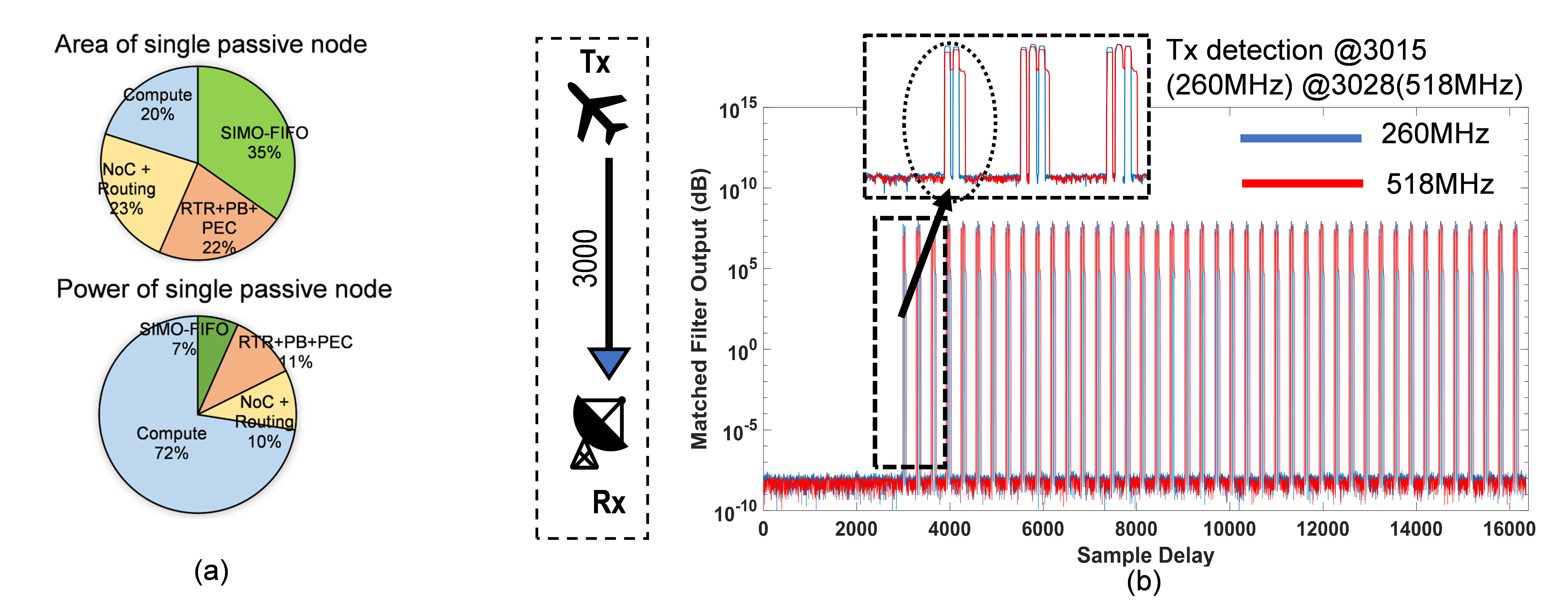}
		\vspace{-0.1in}
	\caption{\textcolor{black}{\small \sl Results of prototype implementation (a)Area and Simulated power breakup (b)Results of Experiment 1 on test-chip}}
	\vspace{-0.05in}
	\label{fig:exp1}
\end{figure*}

\begin{figure*}
	\centering
	\includegraphics[width = 0.9\textwidth]{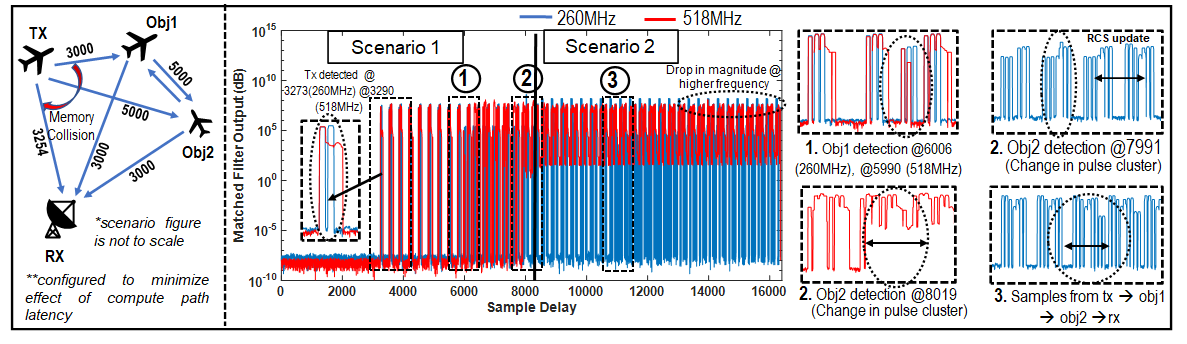}
		\vspace{-0.15in}
	\caption{\textcolor{black}{\small \sl Results of experiment 2 on test-chip}}
	\vspace{-0.1in}
	\label{fig:exp2}
\end{figure*}

\subsection{Measurement Results}
\textcolor{black}{Figure~\ref{fig:chip}(b) shows the experiment setup. A Printed Circuit Board (PCB) is used to apply the stimuli to the packaged testchip. The SCPs are generated in a CPU and loaded into the chip through the SPI using a $0.5$ kbit/s bit-serial interface. This serial interface is used for both loading of scenario configurations and reading output data from the chip. 
The receiver SRAM-based storage circuit on the receiver stores sub-samples of the real-time outputs on-chip. The experiment is repeated multiple times to gather all subsamples required to fully reconstruct the output (offline). This reconstructed output is further post-processed (matched filtered with the input signal) to detect the peaks indicating ranges of objects relative to Rx. }

\textcolor{black}{Figure~\ref{fig:chip}(c) shows the die-shot and test-chip summary. The design achieves up to $518$ MHz of per-channel bandwidth versus $100$ MHz in Colosseum, the state-of-the-art RF emulator \cite{barcklow2019radio,colosseum1} and $180$ MHz in \cite{borries_emulator}. At $518$ MHz, the maximum emulation distance is $9.5$ km (16K samples) with 64 kB SRAM/node and compute latency of a path is $0.24$ $\mu$s. Since the sample storage is implemented with SRAMs, it can only support either read or write in a cycle. The minimum emulation range(sample delay) between two objects is 0.67 km @518 MHz bandwidth (1024 samples in each SRAM plus compute latency for the path). }

\textcolor{black}{Figure~\ref{fig:exp1}(a) shows the area and simulated power breakup of a single node in the test-chip. 
The control path dominates the total area but requires less power than compute modules. The addition of memory collision support increases area and simulated power by $28\%$ and $11.8\%$ respectively. }

\textcolor{black}{For Experiment 1 (Figure~\ref{fig:exp1}(b)), we consider a direct path between the Tx and the Rx with passive objects disabled. Range estimation is performed by matched filtering of samples received at the Rx with the transmitted signal. The measured results show $<$ 1 \% error in range estimation. The appearance of 1st non-zero pulse denotes the range of the Tx from the Rx. 
 We vary the frequency of operation and take measurement results at $260$ MHz and $518$ MHz. Wider detection pulses are observed at higher frequency potentially due to increased supply noise leading to bit errors at the output.}

\textcolor{black}{In the 4-node dynamic experiment (Experiment 2, Figure~\ref{fig:exp2}), we determine ranges of multiple objects interacting in the RF system in the presence of memory collision. Object 1 and the Rx are in memory collision for sample distribution in the Tx node. The range of each object is estimated based on the appearance of new pulses in the matched filtering output. The test-chip is able to separately detect two nodes in memory collision as shown in Figure~\ref{fig:exp2}. The measurement results also demonstrate the impact of reflection between passive objects (new pulses appearing at ~11000 sample delay, Tx–Obj1–Obj2–Rx path delay). Further, this experiment also demonstrates a scenario update, by updating the RCS of passive object 2 in the second scenario (scenario length is much smaller than 1ms in this prototype implementation).  The measured emulated ranges show $0.1-1.1 \%$ error. For these experiments, the coefficients for the fractional delay filters were set to 0 for the positive and negative time lags and to non-zero for the 0-time lag. Additionally, the Doppler frequency was set to 0.}

\textcolor{black}{\section{FPGA Implementation}}

\begin{figure} [t]
\centering
\includegraphics[width=88mm]{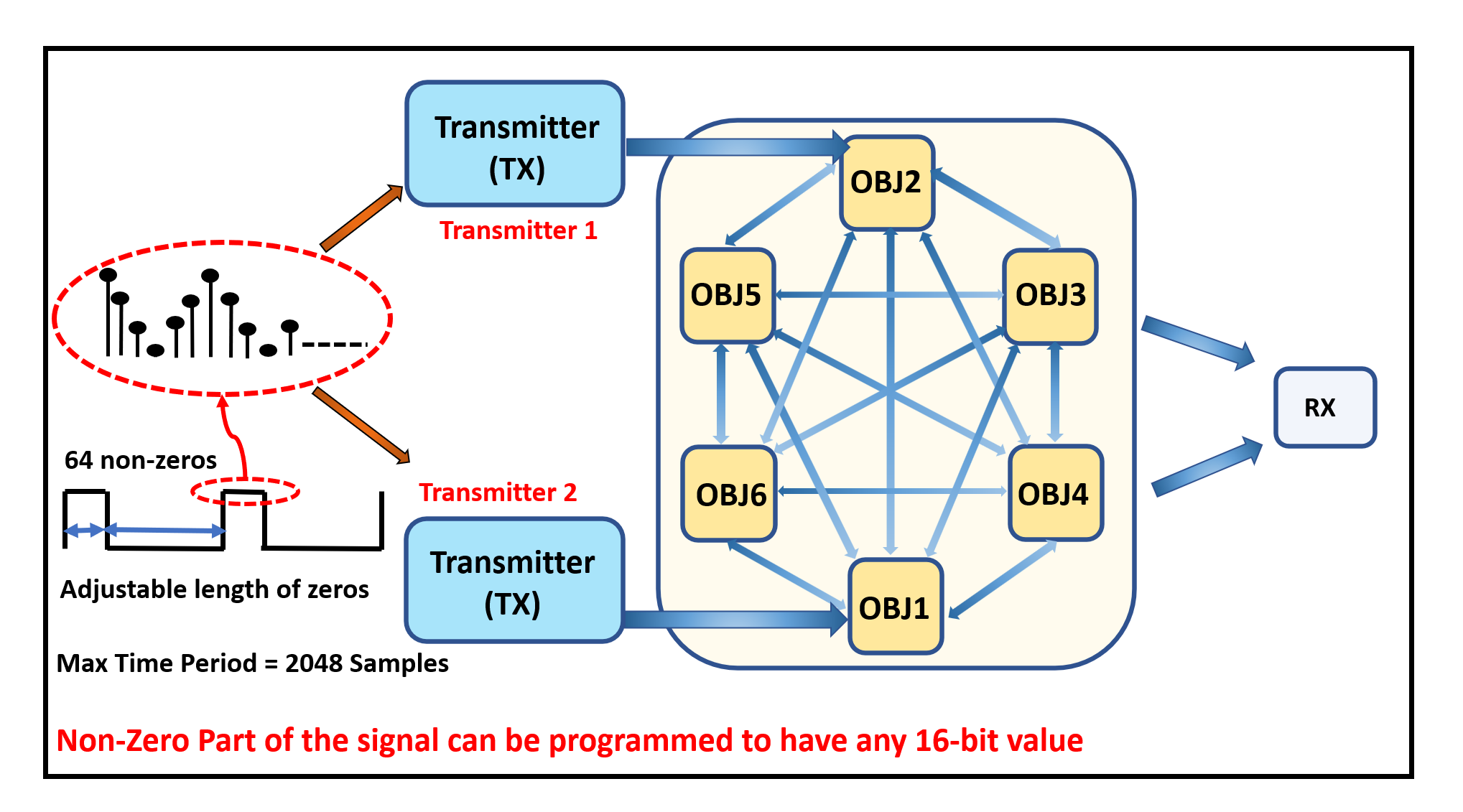}
\caption{The design diagram of the 9-node test case with the signal generator on FPGA.}
\label{fig:diagram}
\end{figure}

\textcolor{black}{The above discussion has showcased the capabilities of our DPCM emulator through ASIC-based tests. In this section, we develop this methodology further to explore more complex scenarios involving a greater number of objects. Utilizing the Xilinx ZCU104 FPGA board, we can enhance our DPCM emulator to simulate the aforementioned larger-scale scenarios, bolstering our claim that the model has promising scalability. In this new paradigm, we emulate a dynamic test environment in a 9-node configuration, which includes 2 Transmitters, 6 Passive interacting reflectors (Object 1 to 6), and 1 Receiver as shown in Figure~\ref{fig:diagram}.}

\subsection{Implementation Details of Data Path}
Our FPGA-based emulator employs the same data path implementation as the chip-test design, as depicted in Figure~\ref{fig:archi}. Each TX consists of a 32-bit Digital Signal Generator (16-bit real and 16-bit imaginary), configurable to generate periodic, digitized RF I/Q samples with a maximum periodicity of 2048 samples. The duty cycle can also be programmed from 3.125\% to 100\% and each non-zero sample can be independently programmed to allow a wide variety of RF patterns to be generated. Then, signals emitted by the TXs are conveyed to the 6 passive objects, where they undergo filtering, attenuation, and delay to generate the reflected signal for each object. The receiver then aggregates these reflected signals from all passive objects, enhances them with a programmable gain, and subsequently, the consolidated output is extracted from the FPGA. The memory used for the receiver is Ultra\_RAM \cite{Xilinx2016UltraRAM} memory provided by the ZCU104 board, and was chosen for data read/write transactions in the receiver. This type of memory supports cascading with an optimized pipeline to enhance timing and reduce latency. However, given the Ultra\_RAM's inability to support a read-only memory (ROM) configuration, ROM is implemented via a Block RAM acting as the lookup table for the Doppler generator.

\begin{figure} [t]
\centering
\includegraphics[width=88mm]{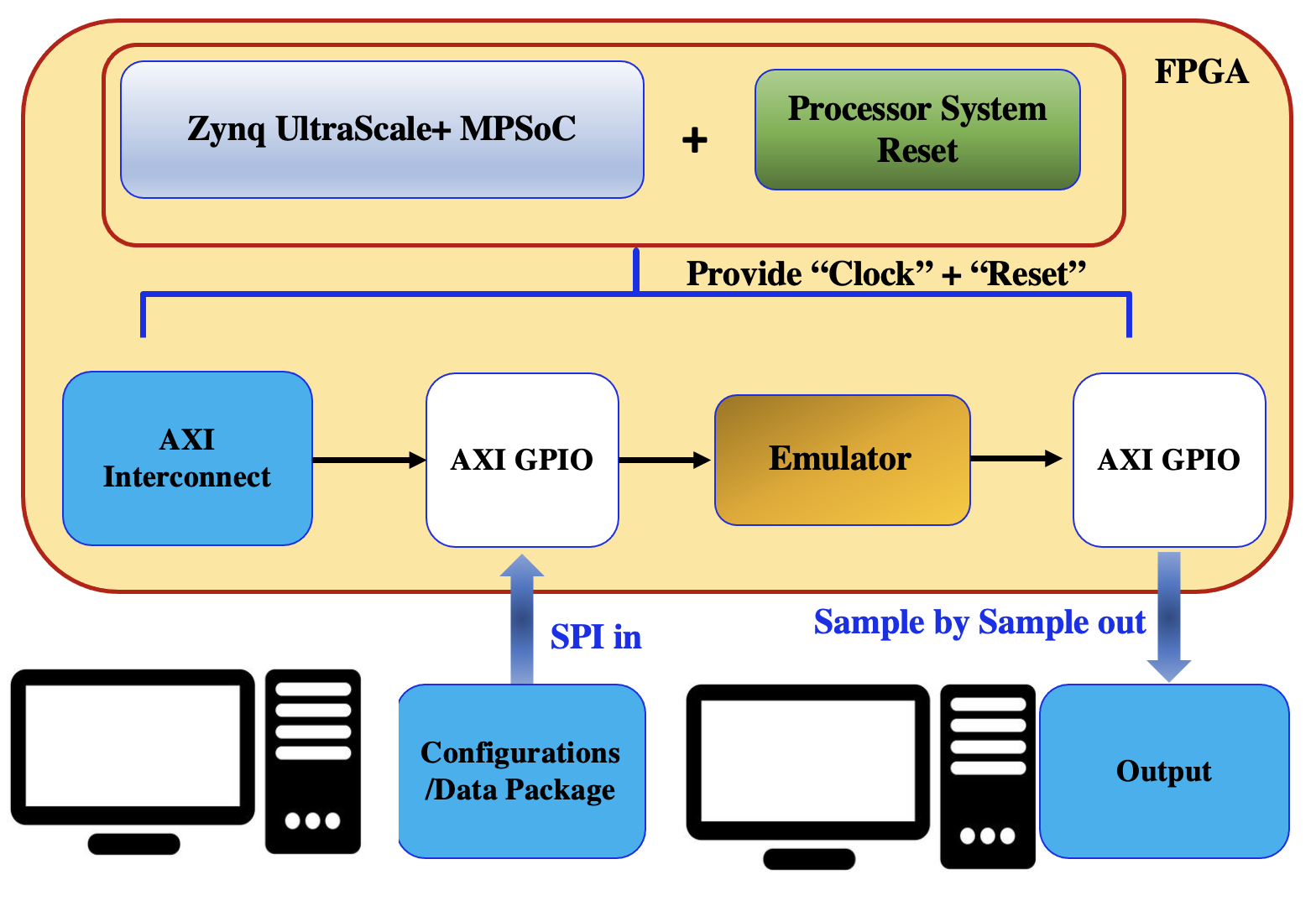}
\caption{\small \sl The Experiment Setup on Xilinx ZCU104 Board.}
\label{fig:FPGA}
\end{figure}

\subsection{Implementation Details of Control Path}
In the FPGA-based design, we employ two distinct variants of Ultra\_RAM memories: Single-port Ultra\_RAM (SURAM) and Dual-port Ultra\_RAM (DURAM), both configured with a depth of 512 samples and a data width of 32 bits. since the memory depth was updated to 512 samples, this necessitated adjustments to the memory collision support, which now offers full support at the updated sample depth. To facilitate simultaneous read/write operations within the same cycle, the RTR leverages DURAM. However, given the finite availability of Ultra\_RAM resources in the ZCU104 board, we have substituted some DURAMs with Dual-port Block RAM (BRAM) to maintain constant performance. Additionally, to meet the high-speed performance requirements and adhere to timing constraints, we have integrated 4 pipeline stages into the critical path of the emulation system's design.

\begin{figure*} [h]
\centering
\includegraphics[width=180mm]{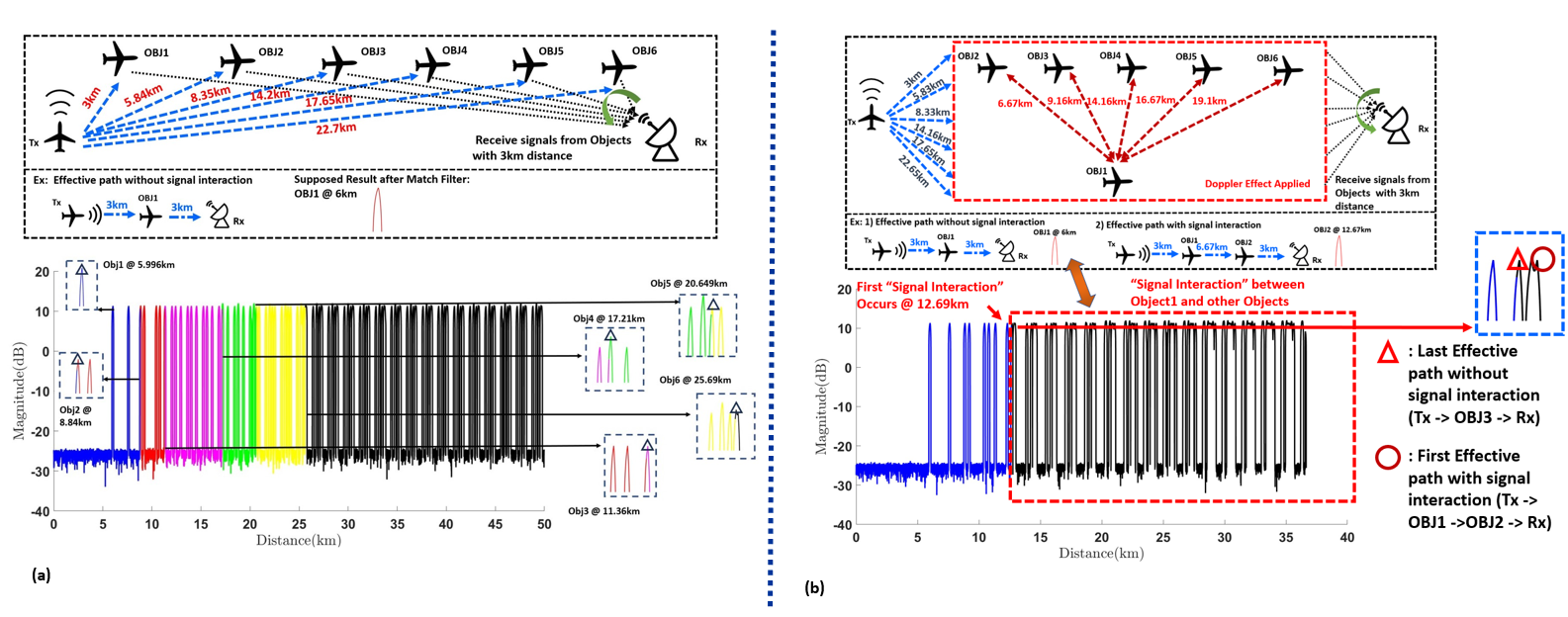}
\caption{\small \sl(a) The experiment and setup of the 9-node test case for the static scenario. (b) The experiment and setup of the 9-node test case for the dynamic scenario.}
\label{fig:8nodes_s_d}
\end{figure*}

\begin{figure*} [t]
\centering
\includegraphics[width=175mm]{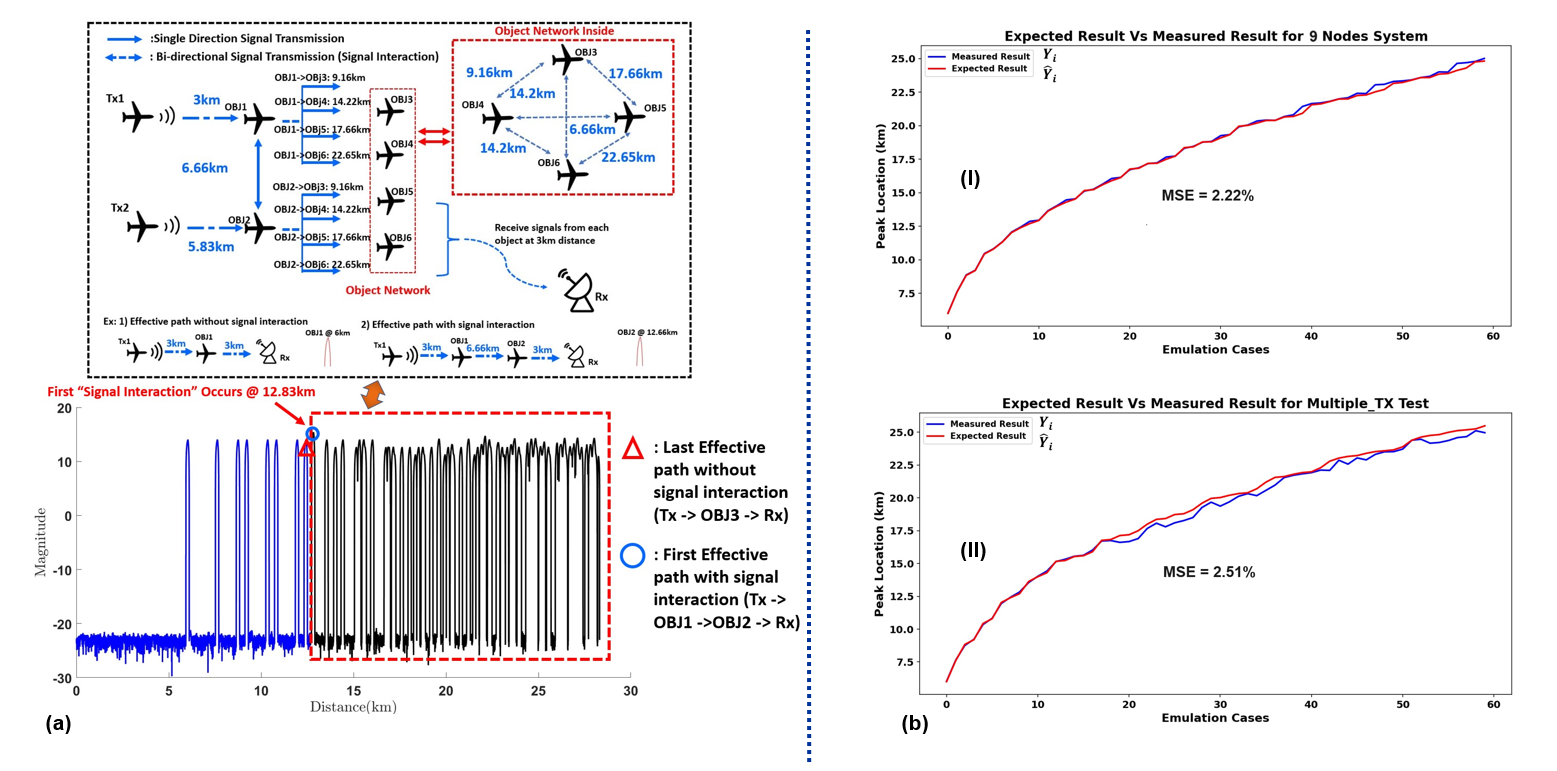}
\caption{\small \sl(a) The experiment result and setup of the multiple TX Test. (b) The error estimation between expected emulation result and measured emulation result: (I) The 9-node test case. (II) The multiple\_tx test case.}
\label{fig:multi_agent_tx_error}
\end{figure*}
\begin{figure} [t]
\centering
\includegraphics[width=85mm]{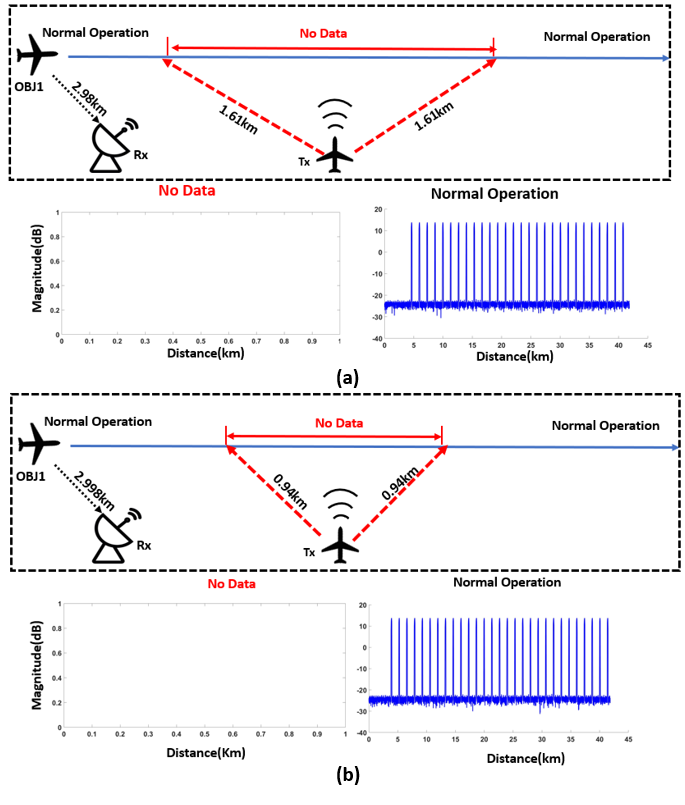}
\caption{\small \sl(a) The experiment result and setup of the Minimum Latency Test with 1024 memory bank. (b) The experiment result and setup of the Minimum Latency Test with 512 memory bank.}
\label{fig:fig:min_latency}
\end{figure}


\subsection{FPGA-based RF Emulator Performace Demonstration}
The measurement setup of the FPGA-based emulator is illustrated as shown in Figure~\ref{fig:FPGA}. A MATLAB-based channel generator is used which, given specific scenario parameters such as object location, distance, RCS, velocity, etc. can generate a configuration packet that is first uploaded from the PC to the FPGA via the IP address of the ZCU104 board, and then is programmed onto the FPGA by reading it through the Advanced Extensible Interface-General Purpose Input/Output (AXI-GPIO) interface. This configuration packet consists of RCS tables, Delay matrices, FIR filter coefficients, and Doppler frequency tables for each respective object to emulate various RF scenarios. The measured output is collected from Xilinx Hardware Management via another AXI-GPIO interface. Finally, we compute the propagation delay of the measured output with respect to the input signal using the MATLAB-based Matched Filter. The performance of the system is evaluated by the accuracy (MSE) of emulation between the expected distance values ($\hat {Y_{i}}$) and the measured emulated distance results ($Y_{i}$) from:
\begin{equation}
\text{MSE} = \frac {1}{n}\sum _{i=1}^{n}(Y_{i} - \hat {Y_{i}})^{2}
\label{eq:MSE}
\end{equation}Several test cases are demonstrated, and details of those are articulated below and all test-cases were run at 215MHz of per-channel bandwidth. 

\subsubsection{Emulation of Range Estimation}
To clearly demonstrate the emulator performance, We first construct a simplified version of the 9-node system to verify the accuracy of range emulation. We activate a single TX and maintain all nodes in a stationary state, simulating only a single reflection per signal. We proceed with the assumption that the passive objects do not reflect signals amongst themselves, effectively disregarding any reflections occurring between objects OBJ1 through OBJ6. Each of the four components is positioned 3 km to 22.7 km from the transmitter, as shown in Figure~\ref{fig:8nodes_s_d}(a). The RX, located 3 km away from each object, accumulates the reflected signals that are sent out of the FPGA. We only enable the RCS to [1, 0, 0, 0, 0, 0] to make sure no reflection signal is getting from other objects. Under this condition, the estimated range of these components exhibits a minor deviation of 0.25\% (Eq.\ref{eq:MSE}) when compared to their actual range.

The second test, shown in Figure~\ref{fig:8nodes_s_d}(b), has the exact initial locations as the first test, except we enable ``inter-object interaction" (IOI). That is, the signal reflected from one object is again reflected by other objects, resulting in a feedback loop (The value of the FIR coefficient remains the same, but the RCS changes from [1, 0, 0, 0, 0, 0] to [1, 1, 1, 1, 1, 1]). Additionally, Object 1 moves toward the other 5 objects with the mutual Doppler effect taken into account. Figure~\ref{fig:8nodes_s_d}(b) shows the measured result of the IOI scenario, where the signal interaction becomes visible after the location is shown in a red dashed line(``Signal Interaction" Occurs). Compared with static cases, under more complex scenarios, each range estimation from the emulation result has a \textless2.22\% error, as indicated by part in Figure~\ref{fig:multi_agent_tx_error}(a.I), compared to the actual range with result illustrated.

The Third test is the emulator with multiple TXs, which is displayed in Figure~\ref{fig:multi_agent_tx_error}(a). Two TXs are applied, and those transfer the RF data to OBJ1 and OBJ2 respectively. Six objects are reflecting the signal within the emulation system, ranging from 6.66 km to 22.65 km. The same as the 9-node test case, the RX keeps at 3 km away from each object. For this test, the final emulation result after the matched filter is less than 2.51\% compared to the actual range (part in Figure~\ref{fig:multi_agent_tx_error}(b.II)).
\begin{figure} [t]
\centering
\includegraphics[width=88mm]{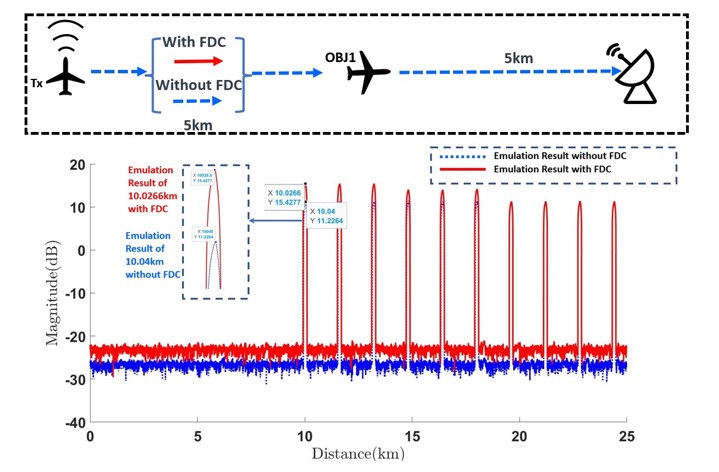}
\caption{The experiment result and setup of the Fraction Delay Correction.}
\label{fig:FDC}
\end{figure}
\begin{figure} [b]
\centering
\includegraphics[width=85mm]{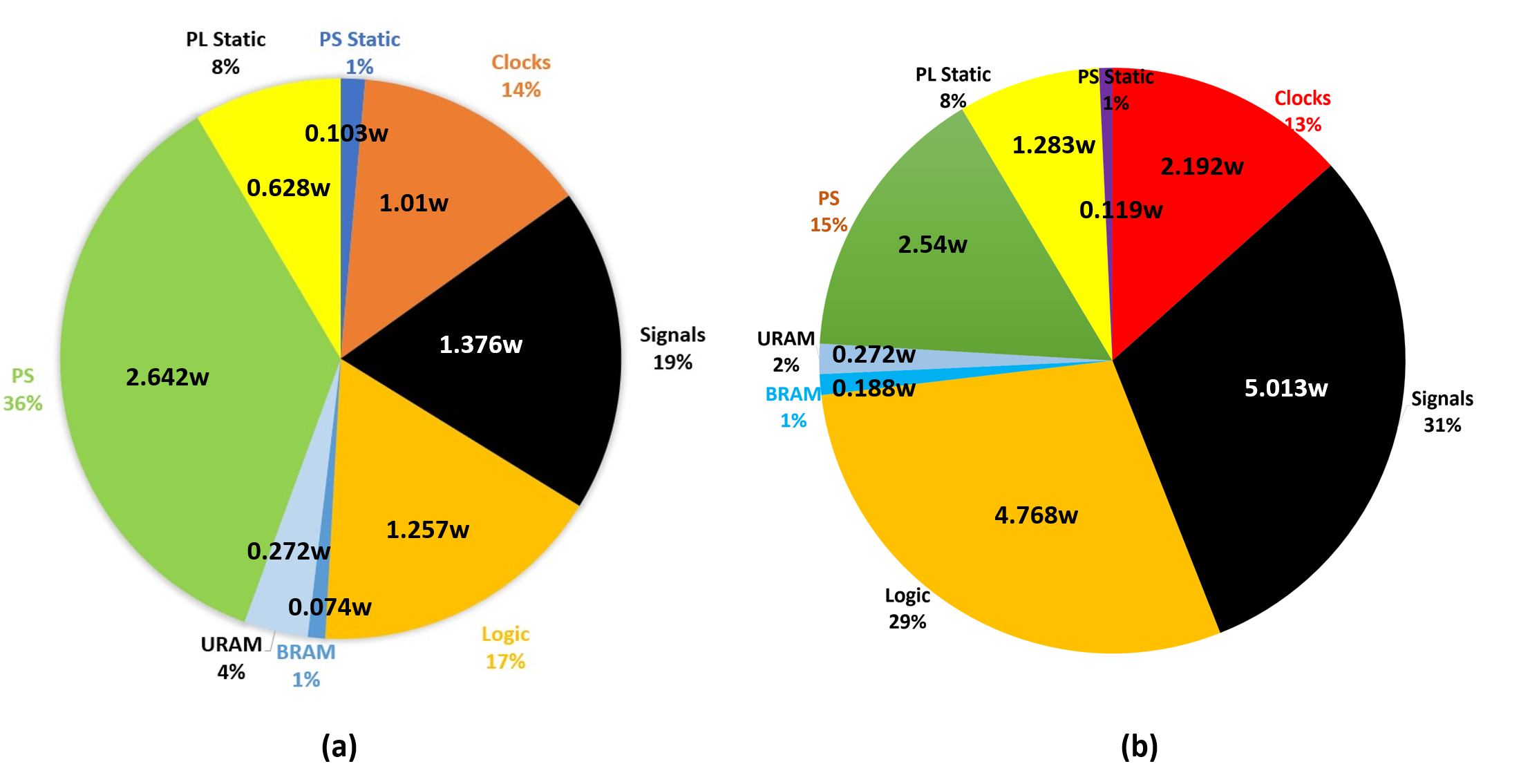}
\caption{(a) Power consumption for each component of the 6-node test case. (b) Power consumption for each component of the 9-node test case. }
\label{fig:power}
\end{figure}

\begin{figure} [h]
\centering
\includegraphics[width=88mm]{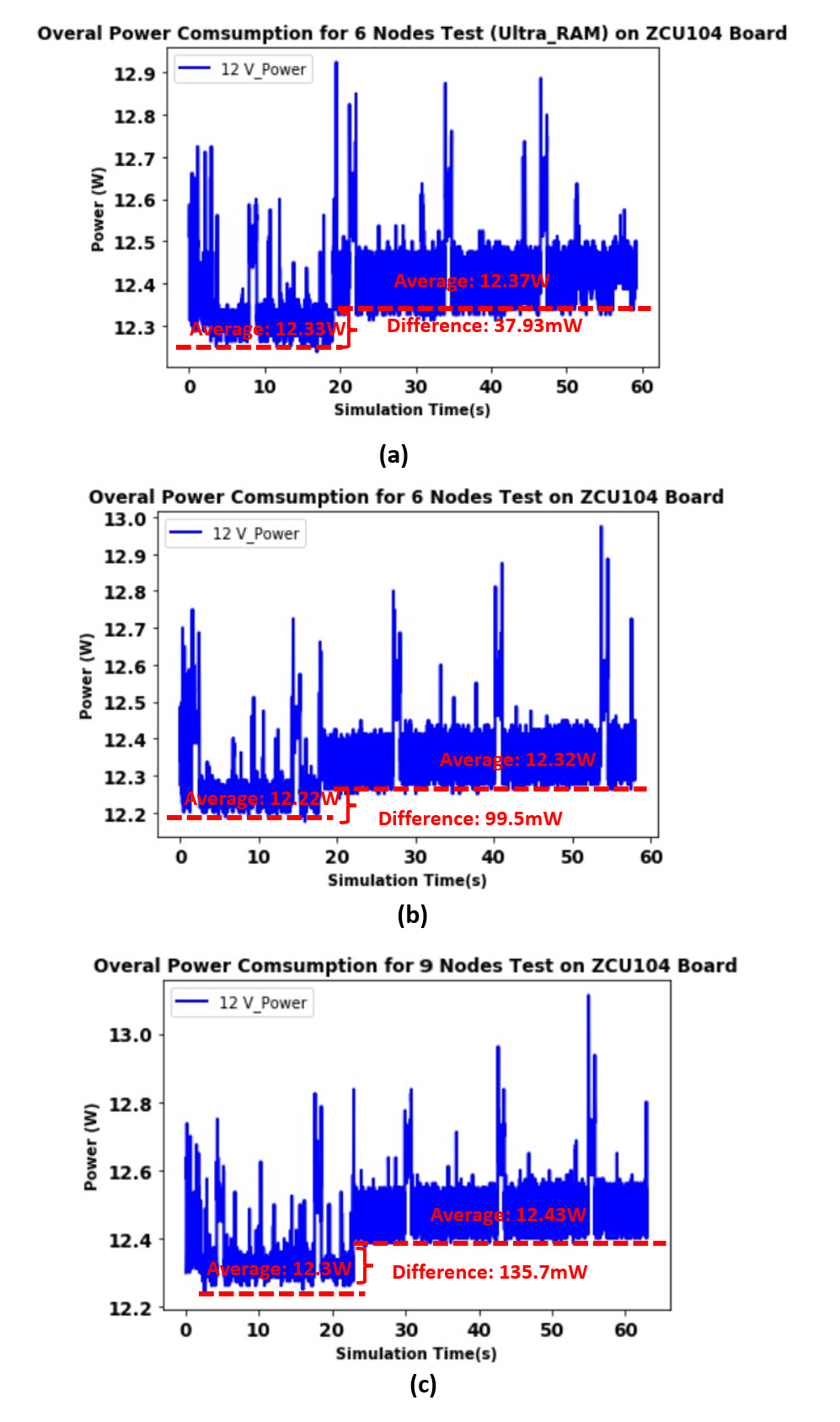}
\caption{Overall Power Consumption on ZCU104 Board: (a) 6-node test case with Ultra\_RAM. (b) 6-node test case with BRAM only. (c) 9-node test case.}
\label{fig:algorithm_power}
\end{figure}

\subsubsection{Minimum Latency Test}
When two objects are too close to each other, the emulator cannot detect the actual range between them because there is a minimum amount of time required by the system to process the collected data (minimum latency of the hardware). Therefore, the test of minimum latency is equivalent to examining the minimum sample range that allows the emulator system to work normally, which is determined by the size of the sub-banks (where simultaneous read and write operations are not possible) and the hardware latency. For this test, we enable only two nodes of the system, TX and Object 1. The TX is located at the center of the origin, and Object 1 is initially located 7.994 km away from the TX. Object 1 begins moving towards the TX in the horizontal direction while TX remains stationary. As shown in Figure~\ref{fig:fig:min_latency}(a), when Object 1 moves to a location within 1.61 km of TX, the signals completely disappear. The ``no signal" status is maintained while Object 1 is within this distance, but as Object 1 moves further than 1.61 km away from TX, the emulator system returns to normal operation. Thus, a minimum range of 1.61 km is required for the system to operate normally. This value is approximately 2.4 times larger than that observed in chip-test case one. \textcolor{black}{The primary reason for this difference, as detailed in Section IV, is related to the formula for buffer delay, which is defined as $\text{buffer delay} = d \times f/c$. Consequently, the actual distance can be calculated as $\text{buffer delay} \times c/f$. The discrepancy between the two test results stems from variations in the frequency ($f$), or bandwidth, used in each test, which ultimately influences the capabilities of the system.}

We repeat the aforementioned experiment using a smaller memory depth of 512 samples. From Figure~\ref{fig:fig:min_latency}(b), the emulator only operates abnormally when Object 1 is located within 0.94 km. It is evident that employing an emulator with lower memory depth effectively reduces the window of read-write collision, resulting in smaller system latency, enhancing (e.g.\ reducing) the minimum emulation range.


{
\setlength{\tabcolsep}{1.25mm}%
\renewcommand{\arraystretch}{1.2}
\newcommand{\CPcolumnonewidth}{25mm}%
\newcommand{\CPcolumntwowidth}{23mm}%
\newcommand{\CPcolumnthreewidth}{23mm}%
\newcommand{\CPcellBold}[1]{\textbf{#1}} 
\newcommand{\CPcell}[1]{#1} 
\begin{table} 
\caption{\small \sl The power, MAX IBW, and utilization of design cells based on 6-node test case and 9-node test case.}
\small
\centering
\begin{tabular}{|l|l|l|l|}\hline
\parbox{\CPcolumnonewidth}{\CPcell{\bfseries Component Name}} & \parbox{\CPcolumntwowidth}{\CPcell{\bfseries 6-node test case}} &
\parbox{\CPcolumnthreewidth}{\CPcell{\bfseries 9-node test case}}\\ 
\hline
\CPcellBold{LUT} & \CPcell{181541} & \CPcell{395450} \\ \hline
\CPcellBold{LUTRAM} & \CPcell{9399} & \CPcell{23081} \\ \hline
\CPcellBold{FF} & \CPcell{243201} & \CPcell{524938}\\ \hline
\CPcellBold{BRAM} & \CPcell{91.5} & \CPcell{259}\\ \hline
\CPcellBold{URAM} & \CPcell{95}  & \CPcell{96}\\ \hline
\CPcellBold{BUFG} & \CPcell{24} & \CPcell{24} \\ \hline
\CPcell{\bfseries MAX IBW} & \CPcell{\bfseries 215 MHz} & \CPcell{\bfseries 215 MHz} \\ \hline
\CPcell{\bfseries Range Estimation} & \CPcell{\bfseries 1.89 km-27.3 km} & \CPcell{\bfseries 1.13 km to 27.3 km} \\ \hline
\CPcell{\bfseries Power On Chip} & \CPcell{\bfseries 37.93 mW} & \CPcell{\bfseries 135.7 mW} \\ \hline
\CPcell{\bfseries Power/100 MHz} & \CPcell{\bfseries 5.58 W} & \CPcell{\bfseries 10.9 W} \\ \hline
\end{tabular}
\label{tab:BRAM}
\end{table}
}

{
\setlength{\tabcolsep}{1.25mm}%
\renewcommand{\arraystretch}{1.2}
\newcommand{\CPcolumnonewidth}{25mm}%
\newcommand{\CPcolumntwowidth}{23mm}%
\newcommand{\CPcolumnthreewidth}{23mm}%
\newcommand{\CPcellBold}[1]{\textbf{#1}} 
\newcommand{\CPcell}[1]{#1} 
\begin{table}
\caption{\small \sl The power, MAX IBW and utilization of design cells based on the 6-node test case.}
\small
\centering
\begin{tabular}{|l|l|l|l|}\hline
\parbox{\CPcolumnonewidth}{\CPcell{\bfseries Component Name}} & \parbox{\CPcolumntwowidth}{\CPcell{\bfseries UltraRAM}} & \parbox{\CPcolumnthreewidth}{\CPcell{\bfseries BRAM only }}\\ \hline
\CPcellBold{LUT} & \CPcell{181541} & \CPcell{180657} \\ \hline
\CPcellBold{LUTRAM} & \CPcell{9399} & \CPcell{9437} \\ \hline
\CPcellBold{FF} & \CPcell{243201} & \CPcell{249804}\\ \hline
\CPcellBold{BRAM} & \CPcell{91.5} & \CPcell{215.5}\\ \hline
\CPcellBold{URAM} & \CPcell{95}  & \CPcell{0}\\ \hline
\CPcellBold{BUFG} & \CPcell{24} & \CPcell{29} \\ \hline
{\CPcell{\bfseries MAX IBW}} & {\CPcell{\bfseries 215 MHz}} & {\CPcell{\bfseries 120 MHz}} \\ \hline
\CPcell{\bfseries Power On Chip} & \CPcell{\bfseries 37.93 mW} & \CPcell{\bfseries 99.5 mW} \\ \hline
{\CPcell{\bfseries Power/100 MHz}} & {\CPcell{\bfseries 5.58 W}} & {\CPcell{\bfseries 5.73 W}} \\ \hline
\end{tabular}
\label{tab:power}
\end{table}
}
{
\setlength{\tabcolsep}{1.25mm}%
\renewcommand{\arraystretch}{1.2}
\newcommand{\CPcolumnonewidth}{20mm}%
\newcommand{\CPcolumntwowidth}{23mm}%
\newcommand{\CPcolumnthreewidth}{23mm}%
\newcommand{\CPcellBold}[1]{\textbf{#1}} 
\newcommand{\CPcell}[1]{#1} %

\begin{table*}
\caption{\small\sl Comparison of our design with prior work.}
\small
\centering
\begin{tabular}{|>{\columncolor[gray]{0.75}}c|c|c|c|c|c|}\hline
\rowcolor{gray!50}\parbox{\CPcolumnonewidth}{\CPcellBold{Attributes}} & \parbox{\CPcolumnonewidth}{\centering\CPcellBold{This Work}} & \parbox{\CPcolumnthreewidth}{\centering\CPcell{D. Barcklow\cite{barcklow2019radio}}} & \parbox{\CPcolumnthreewidth}{\centering\CPcell{Ashish. C \cite{Chaudhari2018}}} & \parbox{\CPcolumnthreewidth}{\centering\CPcell{I. Val \cite{Val2014FPGABased}}} & \parbox{\CPcolumnthreewidth}{\centering\CPcell{kevin. C\cite{borries_emulator}}}\\ \hline
\CPcellBold{Year} & \CPcellBold{Current} & \CPcell{2019} & \CPcell{2018} & \CPcell{2014} & \CPcell{2009}\\ \hline
\CPcellBold{Computation Model} & \CPcellBold{DPCM} & \CPcell{TDL } & \CPcell{TDL} & \CPcell{TDL} & \CPcell{TDL}\\ \hline
\CPcellBold{Channel Tap} & \CPcellBold{4 taps} & \CPcell{4 taps} & \CPcell{5 taps} & \CPcell{-} & \CPcell{3 taps}\\ \hline
\CPcellBold{Max Bandwidth} & \CPcellBold{215 MHz} & \CPcell{80 MHz} & \CPcell{80 MHz} & \CPcell{100 MHz} & \CPcell{90 MHz}\\ \hline
\end{tabular}
\label{tab:comparison}
\end{table*}
}
\subsubsection{Fraction Delay Correction}
In order to highlight the impact of the FDC, we implement a test scenario involving a single object. This object is positioned 5km away from the TX, with the RX set at an equivalent distance from the object (total distance of 10 km from the TX to RX). We measure the emulation result in two distinct conditions: first, with FDC enabled, and second, with FDC disabled. 
According to Figure~\ref{fig:FDC}, the measured distance without FDC (indicated by the Blue Dash Line) after applying the matched filter is 10.04 km.
In contrast, when FDC is enabled in the same scenario, the emulation is able to measure a range of 10.0283 km (Red Solid Line). showing the improved range emulation achieved with FDC.

\subsection{FPGA Hardware Performance}
To further demonstrate the hardware capabilities of our FPGA-based RF emulator, we have designed an additional test case. Specifically, we demonstrate the hardware performance in two scenarios: 1) a 6-node test case featuring a memory depth of 1024, including 1 TX, 4 passive interacting reflectors, and 1 RX; and 2) our illustrated 9-node test case, which extends our evaluation to cover a broader range of interactions and complexities.
\subsubsection{Emulation Time Comparison}
The total time required to run simulations for the 6-node test case, involving four scenarios and each with an 8176-sample test case, is only 0.3612 ms. This is significantly faster compared to the C++ based (10.2521s) or MATLAB (5.2 minutes) based simulator of the physical model. The proposed FPGA system, operating at 215 MHz, significantly accelerates the processing speed and is 28,380$\times$ faster than the C++ Emulator and 82,570$\times$ faster than MATLAB. When the same conditions are applied to the 9-node test case design, the total simulation time is just 0.375ms, while the MATLAB-based simulator needs to take 5.2 minutes (8.32x$10^5$ times longer). As expected, the advantage of emulation is more prominent for test cases with higher complexity such as more objects. 
\subsubsection{Power Analysis}
Figure~\ref{fig:power}(a) and \ref{fig:power}(b) show the total power consumption from Vivado Simulation for both designs on ZCU104 board, which are 7.362 W for the 6-node test case (6.631 W dynamic power consumption and 0.731 W static power consumption) and 16.375W for the 9-node test case design (14.973 W from dynamic power consumption and 1.402 W static power consumption). In Figure~\ref{fig:algorithm_power}, we present a graphical representation of the actual algorithm power consumption recorded from the ZCU104 board for each test case, with a 12 V voltage applied to the power supply. This measurement is conducted using the PMbus library from Python Productivity for Zynq (PYNQ) to monitor the overall power consumption of the ZCU104 board before and after the commencement of the test. Consequently, the algorithm's power consumption from the design is derived as the difference between these two power readings. Figure~\ref{fig:algorithm_power}(a) illustrates the power consumption of the 6-node test case when employing fully the Ultra\_RAM for the design. The graph depicts power measurements taken both before and after system initialization, with values recorded as 12.33 W and 13.37 W, respectively. As a result, the algorithm's power consumption for this particular case is calculated as the difference between these values, equating to 37.93 mW. Similarly, Figure~\ref{fig:algorithm_power}(b) provides the simulation result of the algorithm's power consumption for the same design, albeit with a transition from Ultra\_RAM to BRAM memory, which is roughly 2.6 times larger in amount. Under this scenario, the algorithm's power consumption registers at 99.5 mW. Figure~\ref{fig:algorithm_power}(c) offers the power analysis for a 9-node test case (135.7 mW). Despite the memory size expanding by a factor of 1.8 when transitioning from 6 nodes to 9 nodes, the actual power consumption increases by approximately 3.6 times due to the substitution of certain memory elements from Ultra\_RAM to BRAM.

\subsubsection{Design Component and Bandwidth Analysis}
Table \ref{tab:BRAM} presents the component utilization and total power consumption for both systems. The inclusion of two additional nodes results in a noticeable increase in the memory consumption of the 9-node test case design, approximately three times larger compared to the 6-node test case design. At the same time, the minimum emulation range of the 9-node test case design is broader compared with the 6-node one. Both test cases can achieve a maximum emulated IBW above 215 MHz, which is determined by the max frequency of operation of FPGA. We model the same design by using only BRAM as the memory component without Ultra\_RAM for the 6-node test case. Compared with the Ultra\_RAM version, the IBW is limited to 120MHz. The BRAM-based design also has a higher total power consumption and utilization of design components, which are displayed in Table \ref{tab:power}. 

\subsubsection{Comparison with Prior Works}
Table \ref{tab:comparison} provides a comparative analysis of our design with other state-of-the-art solutions from the past decade. In contrast to existing designs utilizing TDL computation mode, our work is grounded entirely in the DPCM computation model. Notably, we have the capacity to optimize our RF emulation system's maximum bandwidth by employing a 4-tap FIR filter, achieving a remarkable 215 MHz. This figure represents a significant 2.15$\times$ increase over the capabilities of the current single FPGA-based designs.
\section{Conclusion}

{\color{black}

In part I of this series we developed a new ``direct path" computational model for real-time digital RF emulation. We showed, through careful mathematical formulation and simulation, that this model can suitably emulate all channel characteristics necessary for RF system testing. Furthermore, by leveraging mild assumptions on the physical characteristics of scattering profiles and antenna structures our model was shown to yield tremendous computational benefits. This, coupled with a naturally distributed framework, motivated us to explore hardware implementations of the model.

Part II of this series focused on the development of working implementations of the direct path computational model. We approached this from two design perspectives, the first being an ASIC that leverages near-memory computations and autonomous distributed control. This allowed us to achieve high bandwidth and low-latency performance that is not viable in off-the-shelf component-based systems. The second design used an FPGA to implement a larger scale (e.g.\ a greater number of objects) system. Though this style of implementation operates at a lower bandwidth than the ASIC version, it demonstrates that the model can be viably scaled to incorporate more objects as needed.

The complementary results of these two papers have, together, established a new and interesting option for consideration in the future development of RF emulators. Through our collaborative effort, we have coupled innovative modeling techniques with cutting-edge high-performance computing paradigms. The result is a series of implementations that operate at a level of performance not achievable by more traditional designs and a model that can be efficiently scaled to meet testing requirements.

}


 
%

\bibliographystyle{IEEEtran}
\bibliography{IEEEabrv.bib,ref.bib}

\end{document}